\documentclass[journal,twoside,web]{ieeecolor}
\usepackage{generic}
\usepackage{amsmath,amssymb,amsfonts}
\usepackage{graphicx}

\usepackage[absolute]{textpos}
\usepackage{atbegshi}
\AtBeginShipout{%
\begin{textblock*}{.99\paperwidth}[.5,1.5](.5\paperwidth,\paperheight)
\centering\sffamily\scriptsize\noindent
\textcopyright~2024 The Authors. This work is licensed under a Creative Commons Attribution 4.0 License.
For more information, see \href{https://creativecommons.org/licenses/by/4.0/}{https://creativecommons.org/licenses/by/4.0/}.
\end{textblock*}%
}

\usepackage[ruled,vlined,linesnumbered]{algorithm2e}
\usepackage{algpseudocode}

\let\oldnl\nl
\newcommand{\nonl}{\renewcommand{\nl}{\let\nl\oldnl}}

\usepackage[hidelinks]{hyperref}
\usepackage{textcomp}
\usepackage[flushleft]{threeparttable}
\usepackage{booktabs,ragged2e}
\usepackage{mathtools}

\usepackage{amsthm}
\usepackage{mathrsfs}
\usepackage{bbm}
\usepackage{psfrag}
\usepackage{tablefootnote}

\usepackage{makecell}
\usepackage{xspace}

\theoremstyle{definition}
\newtheorem{definition}{Definition}
\newtheorem{theorem}{Theorem}
\newtheorem{corollary}{Corollary}
\newtheorem{remark}{Remark}
\newtheorem{lemma}{Lemma}
\newtheorem{proposition}{Proposition}

\usepackage{accents}

\newcommand{\ubar}[1]{\underaccent{\bar}{#1}}
\newcommand{\minhatA}[2][\hat{A}]{\big[#2\big]_{#1}^+}
\newcommand{\tcite}[1]{\textit{et al.} \cite{#1}}

\newif\ifbuildappendices\buildappendicesfalse


\usepackage[nameinlink]{cleveref}
\crefname{figure}{Fig.}{Figs.}
\crefname{table}{Table}{Tables}
\crefname{lemma}{Lemma}{Lemmas}
\crefname{theorem}{Theorem}{Theorems}
\crefname{algorithm}{Algorithm}{Algorithms}
\crefname{proposition}{Proposition}{Propositions}
\crefname{remark}{Remark}{Remarks}
\crefname{corollary}{Corollary}{Corollaries}
\crefname{appendix}{Appendix}{Appendices}
\crefformat{section}{Section~#2#1#3}

\newcommand{\etal}{\MakeLowercase{\textit{et al.}}\xspace}

\AtBeginShipoutFirst{%
\begin{textblock*}{.84\paperwidth}[.5,0](.5\paperwidth,.235\paperheight)
\centering\sffamily\scriptsize Published in: H.-C. Lin, K.-H. Lin, and H.-Y. Wei, ``Adaptive Age of Information Optimization in Rateless Coding-Based Multicast-Enabled Sensor Networks,'' \emph{IEEE Journal of Selected Areas in Sensors}, vol.~1, pp.~73--92, 2024, doi:~\href{https://doi.org/10.1109/JSAS.2024.3407689}{10.1109/JSAS.2024.3407689}.
\end{textblock*}%
}

\begin{document}
\title{Adaptive Age of Information Optimization in Rateless Coding-Based Multicast-Enabled Sensor Networks}
\author{Hung-Chun Lin,
        Kuang-Hsun Lin,~\IEEEmembership{Member,~IEEE}
        and~Hung-Yu Wei,~\IEEEmembership{Senior Member,~IEEE}
\thanks{This work was supported by Qualcomm under Grant NAT-435534 and the National Science and Technology Council (NSTC), Taiwan, under Grant 112-2628-E-002-025-. \emph{(Corresponding author: Hung-Yu Wei)}}
\thanks{Hung-Chun Lin is with the Graduate Institute of Communication Engineering, National Taiwan University, Taipei City 106319, Taiwan. (e-mail: r10942073@ntu.edu.tw)}
\thanks{Kuang-Hsun Lin is with the Graduate Institute of Communication Engineering, National Taiwan University, Taipei City 106319, Taiwan. (e-mail: f03942140@ntu.edu.tw)}
\thanks{Hung-Yu Wei is with the Graduate Institute of Communication Engineering and the Department of Electrical Engineering, National Taiwan University, Taipei City 106319, Taiwan. (e-mail: hywei@ntu.edu.tw.)}}

\maketitle

\begin{abstract}
As the demand for real-time information in Internet of Things (IoT) and Wireless Sensor Networks (WSN) scenarios grows with the evolution of bandwidth-intensive 5G applications, multicast transmission becomes increasingly vital. This paper delves into the significant role of multicast in WSN, exploring a novel perspective of using rateless codes over the traditionally employed Hybrid automatic repeat request (HARQ) {for eliminating retransmissions} in a wireless multicast system. We aim to optimize data freshness, quantified by the Age of Information (AoI), and analyze strategies to minimize time-average AoI in a multicast environment with diverse channel conditions. {Specifically, our policy focuses on optimizing the time-average AoI based on sensor devices' feedback. We transform the problem into a Markov Decision Process (MDP) to locate optimal and low-complexity suboptimal policies. We present the first age-minimum scheme for rateless code-based wireless multicast systems.} Our numerical simulations reveal that the proposed policies, developed considering unique system structural properties, consistently surpass baseline strategies. We are thus able to preempt updates at the most beneficial time, thereby addressing the issue of the bottleneck device's adverse impact on overall performance.
\end{abstract}


\section{Introduction}
\label{sec:introduction}

\IEEEPARstart{W}{ith} the ubiquitous 5G applications opening up a new era of wireless communication, the demand for bandwidth has rapidly increased. Additionally, in Internet of Things (IoT) and Wireless Sensor Networks (WSN) scenarios, where countless devices ranging from self-driving cars to sensors require real-time information, the need for efficient communication methods is paramount.
{As the most influential mobile communication standard organization, the Third Generation Partnership Project (3GPP) is dedicated to providing solutions that enable IoT and WSN in 3GPP-based systems, such as 5G NR or the upcoming 6G system, while considering the limitations of wireless resources.}
If all receivers need the same information, multicast communication, a transmission strategy that allows packets to be sent from a single sender to multiple receivers simultaneously, is a competitive technology for maintaining freshness and saving bandwidth \cite{Chukhno2023Approaching6G}. In a smart factory, a server multicasts environmental data, like temperature or robot locations, to all robots. These robots act based on this data and their observations, then individually update the server. Similarly, in wildfire detection, positioning algorithms adjust the placement of sensors in real-time, considering factors such as wind direction and vegetation dryness. In high-risk areas, sensors are densely placed, while in low-risk areas, they are more dispersed. The central controller multicasts location directives, saving bandwidth and reducing overhead. In both scenarios, outdated data can lead to severe consequences, from robots' inefficiencies to potential mishaps in other applications. These examples highlight multicast's potential in WSN and the importance of data freshness in real-time applications.

To quantitatively measure the freshness of a device, the metric \textit{Age of Information (AoI)} is proposed in \cite{Kaul2011Minimizing, Kaul2011Onpiggybacking, Kaul2012Realtime} and defined as the elapsed time since the generation of the most recent received update packet for a device. The AoI of the device rises with time and falls when a more recent update packet is obtained. A lower AoI at the device suggests that the data is fresher. Notice that AoI is destination-centric as opposed to packet-centric performance metrics, such as latency.

{The 3GPP, renowned for its contributions to mobile communication standards, plays a crucial role in efficiently utilizing wireless resources and minimizing data redundancy, especially in scenarios like IoT and video streaming. 3GPP has developed comprehensive standards to facilitate multicast communication in various applications. For instance, 3GPP standards encompass group communications for Narrowband IoT (NB-IoT) \cite{3gpp.36.331} and the Multicast Broadcast Service (MBS) \cite{3gpp.23.247}, catering to the specific multicast requirements of these applications.} 
Multicast transmission in highly real-time is also a promising application, as demonstrated by the multicast usage scenario for mission-critical services (MCS) discussed in the 3GPP report \cite{3gpp.23.774}. 3GPP adopts rateless codes as a forward error correction technique to combat transmission error to improve reliability \cite{3gpp.26.502}. The rateless code, or digital fountain code, allows a source to generate an unlimited number of packets from data, and a receiver can decode its data with a sufficient number of packets. {Rateless codes have been applied in the 4G LTE Multimedia Broadcast Multicast Service (MBMS) and extensively studied, demonstrating their effectiveness in improving the reliability of multicast/broadcast transmissions \cite{7866892}. Therefore, 3GPP has also proposed applying rateless codes in 5G NR MBS systems. Rateless codes present a significant advantage in multicast scenarios because they allow the transmitter to skip retransmitting packets with errors and instead send newly generated fountain packets.} 

{Regarding scalability, fixed block-length codes typically require data to be sent at a uniform rate that all receivers can handle. This means the transmission rate must be set based on the receiver with the worst channel conditions (the bottleneck device), leading to inefficiencies as other receivers are underutilized. In contrast, rateless codes allow each receiver to gather packets at their own pace, efficiently addressing the bottleneck issue and enhancing network performance. The LT code \cite{Luby2002LTcodes}, known as the first implemented rateless code, marks a pivotal development in this field. Following them, the Raptor codes \cite{Shokrollahi2006Raptorcode} emerged as a successor, offering significant enhancements in terms of reduced complexity with linear encoding and decoding time. Additionally, the RaptorQ code \cite{rfc6330} has gained popularity as a practical variation of the Raptor code, noted for its improved performance in reducing failure probability and its capacity to support a greater number of source symbols \cite{Raptor2011}.}

Interestingly, attaining the minimum time-average AoI in a multicast network may not imply every delivered status update should be reliable \cite{yates2017timely}. 
{Providing the source sends a new update after all devices decode the current update, the bottleneck device will still affect other devices' AoI performance. 
Therefore, preempting the current update at the right time and generating a new status update may benefit the overall AoI performance.} {Based on the feedback from devices, it is possible to find a policy that minimizes the overall time-average AoI. }
However, it is not trivial to find such a policy due to the complicated combinations of system information, especially as the number of devices in the network increases. Leveraging the knowledge of devices' feedback to develop a more efficient solution is still an open problem.

\begin{figure}
\centering
\includegraphics[width=\linewidth]{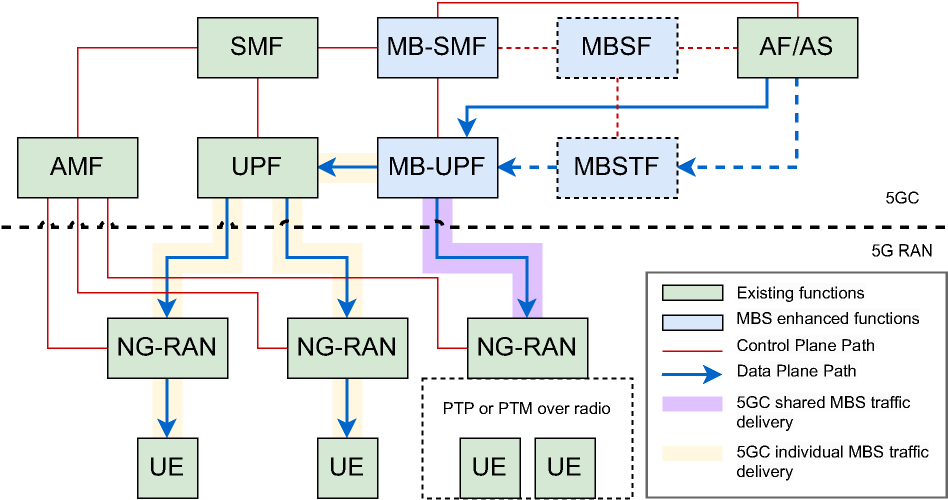}
\caption{5G architecture with MBS enhancement and data transmission diagram\cite{3gpp.23.247}. AMF: Access and Mobility Management Function, AS: Access Stratum.}
\label{5MBS_architecture}
\end{figure}

This paper considers a status update system, where a source multicasts status updates of a physical process to multiple devices. {The status update is encoded via rateless codes, and AoI is adopted as the metric for evaluating the freshness of each device. We aim to find \textbf{the best preemption policy to minimize the time-average AoI by generating a new status update at the appropriate time}.} To our best knowledge, this is the first age-minimum scheme for a rateless code-based wireless multicast status update system. The contributions of this work are the following:
\begin{itemize}
    \item {For a detailed discussion on maintaining a real-time multicast transmission scenario, unlike previous studies using Hybrid Automatic Repeat reQuest (HARQ), we focus on a rateless code-based feedback mechanism to improve data freshness.}
    \item We formulate the system into a communicating Markov decision process (MDP), allowing us to solve the optimal policy by applying the Bellman equation. Further, we exhibit the unique properties of the optimal policy for this MDP, employing these properties to locate the optimal policy.
    \item In the context of large-scale networks, securing the optimal policy becomes impractical due to the curse of dimensionality. As a solution, we explore a low-complexity decentralized algorithm, leveraging structural properties to obtain a suboptimal solution.
    \item We verify the optimal and suboptimal policies' structural properties via numerical simulations. We show that both our proposed policies consistently outperform the baseline strategies proposed in \cite{yates2017timely} with the assistance of AoI information at devices.
\end{itemize}

\section{5G MBS architecture and feedback mechanisms}\label{sec:5MBS_architecture}

There are two delivery modes in 5G MBS: multicast and broadcast. For the broadcast mode, data is transmitted to all User Equipment (UE) in coverage, regardless of Radio Resource Control (RRC) states. However, the broadcast mode fails to manage UEs in the network, and there is no implemented feedback mechanism for the broadcast mode, making it hard to maintain reliability. Hence, in our work, we will focus on the multicast mode. {The multicast delivery mode} {aims to provide high Quality of Service (QoS) services and ensure reliability for a set of registered UEs. We will discuss in more detail how reliability maintenance can be achieved through various feedback mechanisms.} 

The architecture of 5G MBS is listed in the specifications \cite{3gpp.23.247, 3gpp.26.502}, as depicted in \cref{5MBS_architecture}. In addition to the original network functions in the 5G core (5GC) network, there are new network functions defined in the 5G MBS. We list the new functions for 5G MBS enhancement below:

\begin{itemize}
\item \textbf{Multicast/Broadcast Session Management Function (MB-SMF)}: MB-SMF manages sessions, configures MB-UPF for data transport based on policy rules, and optimizes broadcast and multicast MBS sessions with RAN and SMF.
\item \textbf{Multicast/Broadcast User Plane Function (MB-UPF)}: MB-UPF processes packets, enforces QoS and collaborates with MB-SMF for data delivery to RAN nodes or UPF.
\item \textbf{Multicast/Broadcast Service Function (MBSF)}: MBSF bridges MBS and LTE MBMS, manages MBS sessions with AF and MB-SMF, and determines IP multicast addresses.
\item \textbf{Multicast/Broadcast Service Transport Function (MBSTF)}: MBSTF anchors MBS data traffic, sources IP Multicast, and supports related applications, including framing and Forward Error Correction (FEC).
\end{itemize}

For MBS data transmission, there are two possible methods between 5GC and Next generation Radio Access Network (NG-RAN):
\begin{itemize}
    \item \textbf{5GC Individual MBS traffic delivery method}: The single copy of data received by MB-UPF is copied and distributed to individual UE through their dedicated Packet Data Unit (PDU) sessions. Therefore, each such UE requires a PDU session associated with a multicast session.
    \item \textbf{5GC Shared MBS traffic delivery method}: The single copy of data is delivered directly to the NG-RAN, which then distributes it to one or more UEs. The NG-RAN has two possible delivery methods:
    \begin{itemize}
        \item Point-To-Point (PTP) delivery method: Each UE receives a separate copy of the MBS data packet from the RAN node.
        \item Point-To-Multipoint (PTM) delivery method: NG-RAN delivers one copy of the MBS data packets to UEs via radio.
    \end{itemize}

\end{itemize}

While the design of the individual delivery mode seems redundant and less efficient in multicast, it is necessary for those legacy base stations that have not supported MBS yet. Especially when a UE moves from an MBS-capable BS to another that is incapable of MBS, the individual delivery mode is essential to guarantee continuity.

Three key layers ensure reliability for the 5G MBS service: {1) the PHY/MAC layer based on HARQ, 2) the RLC layer based on ARQ, and 3) the Application Layer-Forward Error Correction (AL-FEC).}
For the PHY/MAC layer based on HARQ, the base station identifies the reception status of UEs through HARQ-ACK feedback, which comes in three variants. The first, ACK/NACK-based HARQ feedback, is suitable when the number of UEs is small and assigns a unique Physical Uplink Control Channel (PUCCH) resource to each UE. The second, NACK-only based HARQ-ACK feedback, shares the PUCCH feedback resource among all UEs, and feedback is given only when packet decoding fails. Although it lacks the ability to differentiate between UEs, it can prevent PUCCH overhead. Lastly, no HARQ-ACK feedback can be configured for PTM transmissions with lower reliability requirements to conserve PUCCH resources and reduce latency. The base station can switch between these dynamically, and the UE identifies the active scheme through RRC signaling or a Downlink Control Indicator (DCI). {Whenever the BS detects a NACK for a packet, it triggers retransmission for the failed packet.}

For services that have extreme reliability requirements, PHY/MAC layer HARQ will not be sufficient. Therefore, acknowledged mode (AM) is supported by PTP, which means that ARQ is enabled in the PTP RLC, resulting in high reliability. However, for PTM transmission, only the unacknowledged mode (UM) is available. In addition to the higher design complexity associated with AM for PTM, enabling AM will also cause the performance of UEs with better channel environments to be affected by UEs with poorer environments. 5G MBS base stations can switch between PTP and PTM dynamically in order to meet QoS requirements. This is also a major feature of the 5G MBS standard.

Contrary to the previous two ARQ mechanisms, AL-FEC typically uses a fountain code, such as RaptorQ code, to ensure reliability. Since LTE MBMS, the use of fountain code in multicast has been widely regarded as a relatively efficient choice that can significantly improve performance degradation caused by bottlenecked UEs. In 5G MBS, AL-FEC is mainly implemented in MBSTF. Our work aims to provide a perspective different from traditional HARQ, to explore the possible impact of fountain code on the emerging indicator AoI in 5G MBS, and how to efficiently design its mechanism so that AoI can be optimized.

\section{Related work}\label{related work}

Since the exploration of AoI in a simple First-Come, First-Served queuing model in \cite{Kaul2012Realtime}, most research on AoI has focused on various applications of the queuing systems \cite{AoI_survey}. In addition, AoI analysis on different communication models, e.g., fading wireless channel \cite{Ostman2019fading}, multi-hop wireless network \cite{Talak2021multihop}, and carrier sensing random access (CSMA) protocols \cite{Yates2017multiaccess}, has also been investigated recently. Other notable topics related to AoI include \cite{Zhang2021pricing} in game theory and \cite{Zhang2022AoIvehicular} in information-centric vehicular networks.

Consideration of a multi-user status update system is also a common model. However, it is important to note that not all such systems are multicast. Many systems focus on the scheduling policy of devices to optimize the average Age of Information (AoI). In these cases, they select appropriate devices for transmission among multiple transmitters or receivers. For example, in \cite{Kadota2018broadcast}, Kadota \etal study the optimal policy to minimize AoI in a wireless broadcast network. Hsu \etal \cite{Hsu2020scheduling} consider a random packet arrival system with multi-source and multiple users, and the results are further generalized by considering different queueing disciplines in \cite{Kadota2021MinimizingStochastic}. 
Ceran \etal \cite{Ceran2021ARL} study the optimal scheduling problem with HARQ by different reinforcement learning methods.
{
Zhang \etal \cite{10198349} investigate AoI and PAoI violation probabilities and distributions in an IoT multisource system with preemptive updates. They extend their analysis from single-source systems to account for preemptive disruptions, deriving models for the multisource scenario. Their approach introduces a novel minimization problem focusing on maximal violation probabilities, providing a more comprehensive understanding of timeliness optimization in such systems.
Chen \etal \cite{10078176} first analyze the average PAoI and AoI in a dual update system with sensors having identical service time distributions. They then extend their study to consider sensors with different service time distributions, quantitatively comparing the AoI performance of dual-queue and single-queue systems. Their findings provide practical guidance for communication system design and resource allocation.
Lang \etal \cite{10278911} analyze the average AoI in a remote monitoring system with multiple sensors updating from the same physical process. They employ the SHS method for analysis and find that the logarithm of average AoI decreases linearly with an increase in the logarithm of sensor numbers.
}

In contrast, the source will send a status update to all users in each time slot in the multicast scenario. Tang \etal in \cite{Tang2023BroadcastOrUnicast} evaluate the optimal AoI performance in a multiuser system, comparing the broadcast and unicast transmission schemes across three packet management strategies when random packet arrivals are combined with finite blocklength coding. The remaining studies concerning AoI under multicast can be categorized into two classes. The first class, \cite{zhong2017status,zhong2018multicast,buyukates2019age,chen2020optimizing,Li2020MulticastDeadline}, uses the assumption of shifted exponential service time as the delay time for each device. Each device is guaranteed to be able to receive a status update, while the transmission time differs among all devices, followed by an exponential time distribution with a shifted constant time. A common scheme to determine the transmission time is that the source will transmit a new update as long as $k$ of the $N$ users successfully receive a status update. It is called \textit{earliest k stopping scheme} first proposed in \cite{zhong2017status}. The other class, \cite{hsu2021efficient,nath2018optimum,xie2020age}, focuses on optimizing the trade-off between AoI and energy with different multicast system designs. It can be noticed that recent research rarely considers the effect of the feedback mechanism under multicast on AoI.

{Because the source will decide whether to generate a new update in each time slot adaptively, the number of packets for each status update is not fixed.} The research \cite{yates2017timely,Wei2020Ontheage,Meng2021Spinal,Wang2019whento,Wang2022AgeOptimalTP,Zhou2020minimum,Xie2021online,Feng2019Ageoptimal} focus on nonuniform status update size, but only \cite{yates2017timely} has done a preliminary look under multicast scenario. In \cite{yates2017timely}, Yates \etal analyze the performance of AoI under two coding schemes: finite redundancy (FR) and infinite incremental redundancy (IIR). Unlike the work \cite{yates2017timely}, our paper assumes that devices can give feedback on the reception status of each packet so that the source can know all devices' current packet reception status. We investigate an optimal transmission strategy that can improve AoI more than FR or IIR in this case. In \cite{Wei2020Ontheage}, the peak AoI in a large-scale wireless network using rateless code is analyzed considering sources' distribution. In addition to a different system setting compared to their study, our study focuses on the policy to minimize time-average AoI. In \cite{Meng2021Spinal}, Meng \etal analyze the upper bound of AoI using a new type of rateless code, Spinal codes. The optimal packet preempting policy with non-uniform packet sizes for a Bernoulli source is considered in \cite{Wang2019whento}. Wang \etal \cite{Wang2022AgeOptimalTP} study the impact on two HARQ schemes with a multistate Markov source in space-air-ground-integrated networks.
A joint design of sampling and scheduling policies to minimize AoI with non-uniform packet size is considered in \cite{Zhou2020minimum}, and Xie \etal \cite{Xie2021online} further extend the work by exploiting online policies. However, the work \cite{Zhou2020minimum} and \cite{Xie2021online} concentrate on multiple physical processes and scheduling the best devices to transmit. Instead, we study the best time to preempt the packets under the multicast scenario when we use rateless code with a single process. Feng and Yang \cite{Feng2019Ageoptimal} consider the rateless code in an erasure channel with a single user and aim to find the optimal scheduling policy. Our system further depicts more special properties under the multicast scenario and proposes two practical algorithms as solutions. The closely related work is organized in \cref{related_work}.

\begin{table*}
\begin{threeparttable}
\caption{Comparison of related work}
\label{related_work}
\begin{tabular}{p{2.1cm}p{0.45cm}p{5.9cm}p{2.2cm}p{1.7cm}p{1cm}p{1.9cm}}
\toprule
Paper & Year & Main target & Communication model\tnote{a} & Channel error\tnote{b} & Feedback\tnote{c} & Approach\tnote{d}\\
\toprule
Our work & 2023 & Optimal preemptive policy for multicast with rateless code & 1 to N, multicast & Non-uniform & Yes & MDP\\ 
\midrule
Yates \tcite{yates2017timely} & 2017 & AoI analysis with given transmission schemes & 1 to N, multicast & Uniform & Limited & Probability\\
Tang \tcite{Tang2023BroadcastOrUnicast} & 2023 & AoI performance using broadcast and unicast schemes with finite blocklength code  & 1 to N, multicast and unicast & Non-uniform & No & Probability\\
Zhong \tcite{zhong2017status} & 2017 & Optimal transmission time in a random delay network & 1 to N, multicast & No & Yes & Probability\\
Li \tcite{Li2020MulticastDeadline} & 2020 & AoI and peak-AoI for earliest stopping k scheme subject to deadlines & 1 to N, multicast & No & Yes & Probability\\
Chen \tcite{chen2020optimizing} & 2020 & Optimal transmission interval without feedback & 1 to N, multicast & No & No & Probability\\
\midrule
Feng \tcite{Feng2019Ageoptimal} & 2019 & Optimal transmission scheme with Rateless code & 1 to 1, unicast & Yes & Yes & MDP\\
Wang \tcite{Wang2022AgeOptimalTP} & 2022 & Optimal transmission scheme for classical and IR HARQ with a multistate Markov source & 1 to 1, unicast & Yes & Yes & MDP\\
Wang \tcite{Wang2019whento} & 2019 & Optimal preemptive policy with packets of non-uniform sizes and random arrivals& 1 to 1, unicast & No & No & MDP\\
Kadota \tcite{Kadota2018broadcast} & 2018 & Performance comparison of scheduling policies & 1 to N, unicast & Non-uniform & Yes & DP, LO, RMAB \\
Ceran \tcite{Ceran2021ARL} & 2021 & Optimal scheduling policy with HARQ & 1 to N, unicast & Non-uniform & Yes & RL\\
Wei \tcite{Wei2020Ontheage} & 2020 & Peak-AoI with given coding schemes and networks & N to 1, unicast & Yes & Yes & Probability \\
Zhou \tcite{Zhou2020minimum} & 2020 & Optimal scheduling and sampling policy with different packet sizes & N to 1, unicast & Non-uniform & Yes & MDP\\
Hsu \tcite{Hsu2020scheduling} & 2020 & Optimal scheduling policy with multiple sources & N to N, unicast & No & No & MDP\\
\bottomrule
\end{tabular}
\smallskip
\scriptsize
\begin{tablenotes}
\RaggedRight
\item[a] 1 to N: There are one transmitter and multiple receivers; 
unicast: The packet is transmitted to only one receiver in each time slot;
multicast: A packet is transmitted to all receivers
\item[b]No: A packet will not be lost in the channel;
        Yes: There is only one receiver, and a packet may fail to be delivered due to channel noise; 
        Uniform: All receivers have the same error probability;
        Non-uniform: The error probabilities can be different among all receivers
\item[c] Limited: Feedback is only available after a full status update has been received; 
         Yes: A receiver gives feedback every time when it receives a packet.
\item[d] DP: Dynamic Programming; 
         LO: Lyanopov Optimization; 
         RMAB: Rateless Multi-Armed Bandit;
         RL: Reinforcement Learning
\end{tablenotes}
\end{threeparttable}
\end{table*}

\section{System model and problem formulation}\label{system model}

{In this section, we {begin} by outlining what we mean by a wireless multicast status update system in our study. Following that, we {provide} a comprehensive mathematical explanation of this system. We will also cover the system's state, including the AoI for both the status updates and the sensor devices, as well as the remaining packets that need to be received. Lastly, we model the system's behavior and establish it as a Markov Decision Process.}

\begin{figure}
\centering
\includegraphics[width=\linewidth]{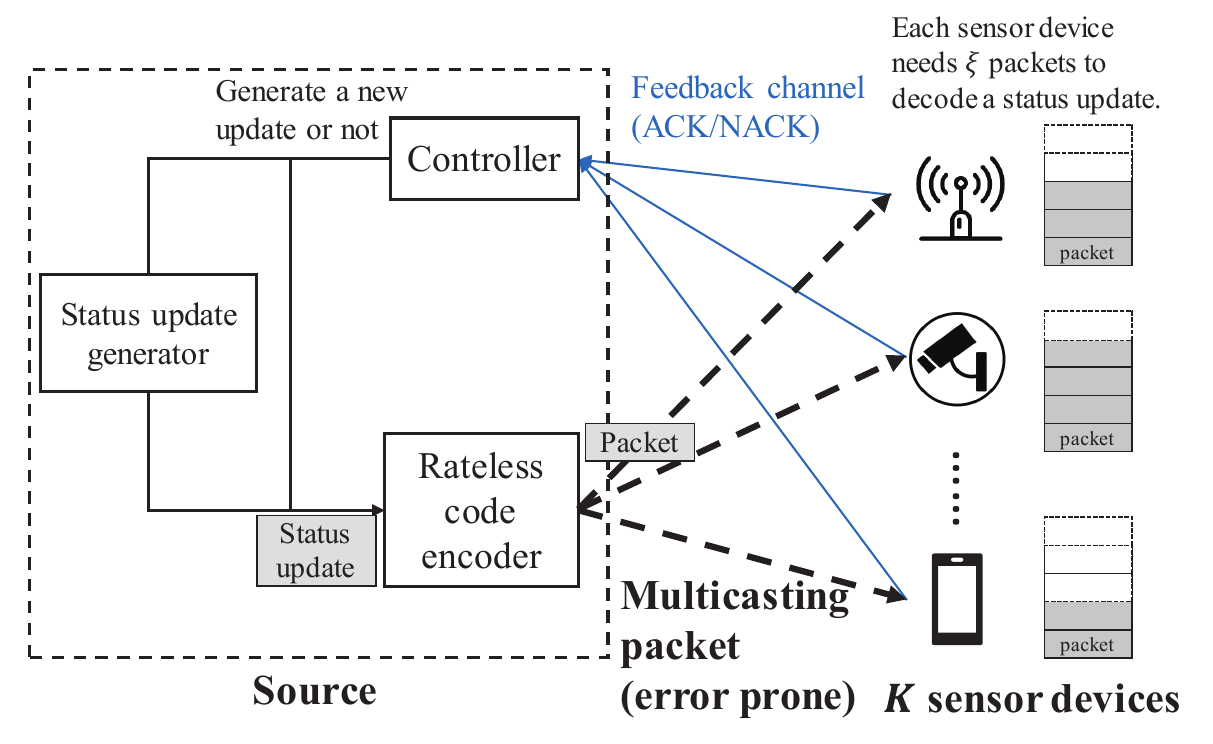}
\caption{A multicast-enabled wireless sensor network with rateless code.}
\label{system_architecture}
\end{figure}

\subsection{Wireless multicast status update system}

We consider a multicast-enabled wireless sensor network consisting of a source and $K$ sensor devices, as shown in \cref{system_architecture}. The source sends status updates (e.g., humidity, temperature, location, etc.) to all other devices in the network via multicast. Since wireless transmission is susceptible to noise and interference, we use rateless code to protect status updates according to 3GPP recommendation \cite{3gpp.26.502} for multicast networks. {Our assumption of rateless code is based on the practically implemented rateless code, RaptorQ code \cite{rfc6330}. We assume the encoder can produce any number of encoded symbols on the fly from the status update as needed.} {Each symbol will have the same size and be transmitted as a MBS packet.} {In this discussion, it is imperative to distinguish between two key terms: a \emph{status update} and \emph{packets}. {A \emph{status update} refers to a complete snapshot of information that needs to be communicated from the source to the receiver. Each snapshot is then protected by rateless codes and transformed into smaller encoded units termed \emph{packets}.}} {{It is assumed that each device is able to decode a status update after successfully receiving $\xi$ packets.} This uniform requirement of $\xi$ packets for all devices is due to the nature of rateless codes. These codes allow for efficient decoding with a high probability upon receiving a slightly larger number of packets than the minimum required. Since all sensors request the same update and the decoding capability is not sensor-specific, $\xi$ remains constant across different devices.} 

{Each device sends an ACK or NACK through a perfect feedback channel, similar to the feedback channel of the HARQ process, to the BS after each packet is received. However, no feedback is needed whenever a status update is successfully decoded. This is to inform the source whether the packet was received successfully, but no retransmission will be triggered for NACK feedback.} The controller in the source {, guided by feedback from all devices, adaptively determines} whether the next transmission should {persist with the current} status update so that more devices can decode correctly or whether it should generate a new status update but discard all packets {associated with the previous} status update. Our goal is to find the optimal strategy for the controller to minimize the sum of time-average AoI for all devices.

\subsection{Mathematical formulation of system model}
We consider a wireless multicast network consisting of one source and a set $\mathcal{K} = \{1, 2, \cdots, K\}$ of $K$ devices. {All of these devices are keen to receive status updates.} Let the time be slotted, indexed by $t = 1,2,\cdots$, and the source can only transmit one packet in a time slot. {Every sensor device can receive this packet within the duration of a time slot and will update its information when the slot concludes.}{\footnote{RaptorQ code has been shown to possess relatively low encoding and decoding complexity, which is shown to be linear. Therefore, we do not consider the decoding time in our work.}}. Consider a binary erasure channel (BEC). For device $k\in\mathcal{K}$, we assume it may fail to receive a packet with probability $p_k \in [0,1)$, defined as the packet erasure rate of device $k$ in our system. While the probability differs between devices, it does not vary with time. Let $u_k(t)\in\{0,1\}$ denotes whether the device $k$ receive a packet successfully in time slot $t$, where $u_k(t) = 1$ with probability $1-p_k$ and $u_k(t) = 0$ with probability $p_k$.
{Each device cannot decode the status update until it successfully receives $\xi$ packets, regardless of their order. The assumption of a rateless code led us to propose $\xi \geq 2$ in our work.}
{The arrival model is a \emph{generate-at-will} model (i.e., the source can generate a message at any time).
{As considered in \cite{Kadota2018broadcast, Li2020MulticastDeadline, Zhou2020minimum}, we assume each} device will give the source an ACK or NACK message via an error-free and instantaneous feedback channel.}
{They do not have to report other information because the source can precisely track each device's status, as we will see in \cref{sec:AoI_transition}. This can help to reduce the uplink feedback overhead.}

Upon receiving the feedback at the beginning of each time slot, the source {is faced with a decision regarding the packet to transmit.} {The decision, denoted by $d(t)$, is about whether to \textit{continue transmitting packets from the current status update} or to \textit{commence with a new status update's packet}. Specifically, $d(t)=0$ indicates the transmission of a packet from the ongoing status update, whereas $d(t)=1$ signals the initiation of a new status update's packet transmission.} {{Initiating} a packet transmission from the new status update will {render} all remaining packets of the ongoing update outdated, prompting their discard across all devices.} Since perfect feedback exists at the end of each time slot, the source can know the exact AoI at each device and how many packets each device still needs to decode the status update. This paper aims to \textit{find the best action for the source according to the current state, and thus, the time-average AoI performance is minimized.} 
 
{At the start of time slot $t$, we define the number of transmitted packets of the latest generated status update as $P(t)$, which {represents} the age of the status update.} In addition, we define the age at the destination device $k$ as $A_k(t)$ for each device. 
For $P(t)$ and $A_k(t)$, we further define their upper limit as $\hat{A}$. 
{This approach is supported by the methodology adopted in related studies \cite{Hsu2020scheduling, Zhou2020minimum}, where finite upper limits are set to manage the computational feasibility of solving average-cost {MDPs} with potentially infinite states. Additionally, this strategy addresses practical considerations; updates beyond a certain age threshold, $\hat{A}$, {become extremely out-of-date and offer similar utility for the devices} \cite{Gindullina2021AoIupper_stale, Leng2019AoIupper_stale}. Thus, imposing an upper limit on age, which may be considered large yet finite, ensures both computational tractability and practical utility.}
The state space of $P(t)$ and $A_k(t)$ are $\mathcal{P}\triangleq \{1,...,\hat{A}\}$ and $\mathcal{A} \triangleq \{\xi,\xi+1,...,\hat{A}\}$, respectively. A device has to receive $\xi$ packets to decode a status update, so its age will not be smaller than $\xi$. Thus, the smallest possible age of a destination will be $\xi$.
$R_k(t) \in \mathcal{R} \triangleq \{0,1,...,\xi\}$ is used to denote the number of packets still needed for device $k$ to decode the update at the beginning of the time slot $t$. If $R_k(t)=0$, the device $k$ has successfully decoded the most recent message, and there is no more packet required for device $k$ before the advent of the next update.

\begin{table}\centering
\caption{List of main notations}
\label{main_notations}
\begin{tabular}{p{1cm}p{6cm}}
 \toprule
Notation & Description \\
 \toprule
$\mathcal{K}$ & Set of devices with the cardinality $|\mathcal{K}|=K$ \\
$\xi$  & Number of packets needed to decode a status update\\
$p_k$& Probability that device $k$ fails to receive a packet\\
$u_k(t)$ & Indicator showing whether device $k$ successfully receives a packet at time slot $t$\\
$d(t)$ & Action that the source takes in time slot $t$ \\
$\mathcal{D}$ & Set of available actions for the source\\
$P(t)$ & Number of packets has been generated by the status update at the beginning of time slot $t$.\\
$\mathcal{P}$ & State space of $P(t)$\\
$A_k(t)$ & Age of device $k$ at the beginning of time slot $t$\\
$\mathcal{A}$  & State space of $A_k(t)$\\
$\hat{A}$ & Upper limit of the AoI\\
$R_k(t)$ & Number of packets still needed for device $k$ to decode a update at the beginning of time slot $t$\\
$\mathcal{R}$ & State space of $R(t)$\\
\midrule
$X_k(t)$ & System state vector of device $k$ at slot $t$\\
$\mathcal{X}_k$ & System state space of device $k$\\
$X(t)$ & System state vector at time slot $t$\\
$\mathcal{X}$ & System state space\\
\midrule
$\Delta$ & Markov decision process for the system\\
$\Pi$ & Set of stationary deterministic policies\\
$\pi$ & Stationary deterministic policy\\
$V^\pi(X)$ & Long-term time-average AoI under policy $\pi$ with initial state $X$\\
$\pi^\star$ & Age-optimal policy\\
$h(\cdot)$ & Relative value function \\
$h_n(\cdot)$ & Relative value function at iteration $n$\\
$Q(\cdot,\cdot)$ & Q-factor\\
$Q_n(\cdot,\cdot)$ & Q-factor at iteration $n$\\
$\mathbf{A}$ & Vector of all $A_k$\\
$\mathbf{R}$ & Vector of all $R_k$\\
$\pi^\star_n$ & Age-optimal policy at iteration $n$\\
\midrule
$\widetilde{X}_k$ &Extended state vector of device $k$\\
$\widetilde{\mathcal{X}}_k$ & Extended state space of device $k$\\
$\Delta_k$ & Markov decision process for device $k$\\
$\widetilde{\pi}^\star$ & Suboptimal policy \\
$h_k(\cdot)$ & Relative value function for $\Delta_k$\\
$h_k^n(\cdot)$ & Relative value function at iteration $n$ for $\Delta_k$\\
$Q_k(\cdot,\cdot)$ & Q-factor for $\Delta_k$\\
$Q_k^n(\cdot,\cdot)$ & Q-factor at iteration $n$ for $\Delta_k$\\
\bottomrule
\end{tabular}
\end{table}

\subsection{Age of Information transition}\label{sec:AoI_transition}
\begin{figure}
\centering
\includegraphics[width=\linewidth]{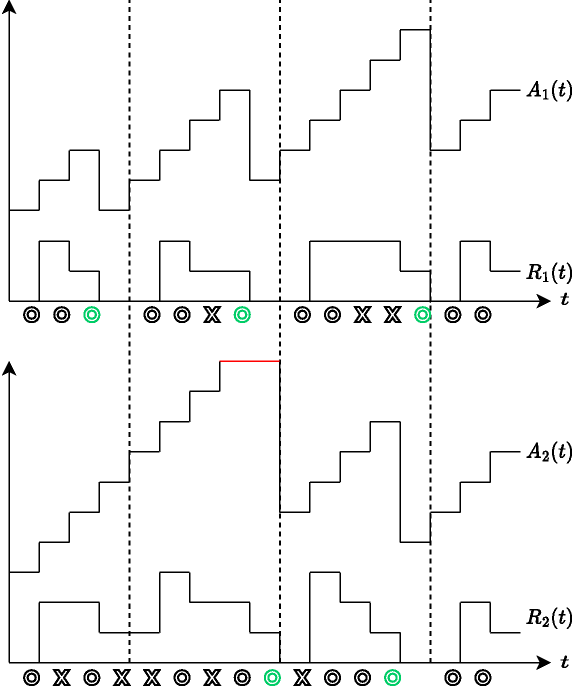}
\caption{An example illustration with system parameters $K=2$, $\hat{A}=10$ and $\xi=3$. Both $A_k$ and $R_k$ are right-continuous. The circle and cross symbols indicate the ACK/NACK feedback. The green circles imply a device has decoded a status update, and thus AoI decreases. The red line means the upper limit of AoI. The source chooses to transmit a new update's packet at time slots $t=1,5,10,15$ in this example.}
\label{toy_example}
\end{figure}
By $X_k(t) \triangleq (A_{k}(t),R_k(t)) \in \mathcal{X}_{k}$, we define the system state vector of device $k$ at slot $t$, and $\mathcal{X}_{k} \subset \mathcal{A} \times \mathcal{R}$ as the system state space of device $k$. Let $X(t) \triangleq (P(t),(X_k(t))_{k\in\mathcal{K}}) \in \mathcal{X}$ be the system state vector at slot $t$, and $\mathcal{X} \subset \mathcal{P} \times \prod_{k\in\mathcal{K}}\mathcal{X}_{k}$ is the system state space. Note that not all combinations of $P(t), X_k(t)$ belong to $\mathcal{X}$. We will discuss the formal definition of the system state space in \cref{Formal definition of the state space}.

If $d(t)=1$, the system will transform from the state $X(t) = (P(t),(X_k(t))_{k\in\mathcal{K}}) $ to $X(t+1) = (P(t+1),(X_k(t+1))_{k\in\mathcal{K}})$ by the following equations:
\begin{equation}\label{AoI_model_d1_P}
    P(t+1) = 1;
\end{equation}
\begin{equation}\label{AoI_model_d1_Ak}
    A_k(t+1) = \minhatA{A_k(t)+1};
\end{equation}
\begin{equation} \label{AoI_model_d1_Rk}
R_k(t+1)=
\begin{cases}
\xi-1&\text{if $u_k(t) = 1$;}\\
\xi&\text{if $u_k(t) = 0$,}
\end{cases}
\end{equation}
where $\minhatA{n} \triangleq \min\{n,\hat{A}\}$ to ensure the age will not surpass the upper limit of AoI. {Since the source selects to transmit the packet from a newly generated update, the age of the status update will reduce to one at the beginning of the next time slot.} The age of the destination will increase by one because it cannot decode an update in the current time slot due to the assumption of $\xi \ge2$. If the device $k$ fails to receive the packet, the number of packages it needs to decode the status update, $R_k(t)$, will remain $\xi$. Otherwise, it will reduce to $\xi-1$.

On the other hand, if $d(t)=0$, the system will transform from the state $X(t)$ to $X(t+1)$ by the following equations:
\begin{equation}\label{AoI_model_d0_P}
    P(t+1) = \minhatA{P(t)+1};
\end{equation}

\begin{equation}\label{AoI_model_d0_Xk}
X_k(t+1)=
\begin{dcases}
\mathmakebox[10em][l]{(\minhatA{A_k(t)+1},R_k(t))}
\text{if $u_k(t) = 0$;}\\[1ex]
(\minhatA{A_k(t)+1},\max\{R_k(t)-1,0\})\\
\qquad\text{if $u_k(t) = 1$ and $R_k(t)\neq 1$;}\\[1ex]
(\minhatA{P(t)+1},0)\\
\qquad\text{if $u_k(t) = 1$ and $R_k(t) = 1$.}\\
\end{dcases}
\end{equation}

{Because the source decides to {continue} transmitting the packet from the previous update, the age of the message, $P(t)$, will increase by one.} If the device $k$ fails to receive the packet, its age will increase, and the number of packets needed to decode the status update remains. {In the case of successfully receiving a packet, the device's age may be updated if it only needs one packet to decode the status update in this time slot ($R_k(t) = 1$).} Interestingly, the age at the destination is the same as the number of packets that have been transmitted after successfully decoding a message. Conversely, if device $k$ cannot decode a status update in the time slot ($R_k(t) \neq 1$), its age increases by one, and the number of packets {required} for decoding reduces by one. From the definition of $R_k(t+1)$, it will never be negative. Therefore, we take a maximum of zero and $R_k(t)-1$. \cref{toy_example} illustrates an example of state transitions to help envision the above transition models. The key notations are listed in \cref{main_notations}.

\subsection{Markov decision process}
Based on the age of information model, we can formulate our problem as a Markov decision process $\Delta$ by the following components \cite{Puterman}:
\begin{itemize}
    \item \textbf{States}: At the beginning of each time slot $t$, we define the system state vector as 
    \begin{equation}
    \begin{aligned}
        X(t) &= (P(t),X_k(t)_{k\in\mathcal{K}}) \\
        &= (P(t),(A_k(t),R_k(t))_{k\in\mathcal{K}}),  
    \end{aligned}
    \end{equation}
    where $P(t)$ is the number of packets transmitted for the latest update, and $X_k(t)$ is the state of the destination. $X_k(t)$ can be decomposed to $A_k(t)$ and $R_k(t)$, representing the age and the number of required packets at the destination, respectively. $\mathcal{X}$ is used to denote the state space, which is finite and bounded in our model.
    
    \item \textbf{Actions}: At each time slot $t$, the source should select an action $d(t)\in\mathcal{D}=\{0,1\}$, depending on the current state $X(t)$.
    
    \item \textbf{Transition probabilities}: For simplicity, we omit the time slot $t$. By taking the action $d$, the transition probabilities from the state $X=(P,(X_k)_{k\in\mathcal{K}})$ to $X'=(P',(X'_k)_{k\in\mathcal{K}})$ can be written as the equation below:
    \begin{equation}\label{state transition probabilities}
        Pr[X'|X,d]=\prod_{k\in\mathcal{K}}Pr[X_k'|X_k,d].
    \end{equation}
    From \eqref{AoI_model_d1_P} and \eqref{AoI_model_d0_P}, the transition of $P$ is determined once the action $d$ is selected. Therefore, it will not affect the transition probabilities, and we only need to consider the uncertainty of $X_k$. For each system state of device $k$, we have
    \begin{equation}\label{Transition prob in ori}
        \begin{aligned}
        Pr[X_k'|X_k,d]=
            \begin{cases}
            1-p_k&\text{ if $X_k'=X_{k,s}$;}\\
            p_k&\text{ if $X_k'=X_{k,f}$;}\\
            0&\text{ otherwise.}
            \end{cases}
        \end{aligned}
    \end{equation}
    
    Here, $X_{k,s}$ denotes the state {where} device $k$ succeeds in receiving the packet, or it will transit to the state failing to get the packet, $X_{k,f}$. According to the \eqref{AoI_model_d1_Ak}, \eqref{AoI_model_d1_Rk} and \eqref{AoI_model_d0_Xk}, we can describe them as
    \begin{equation}
    X_{k,s}=
    \begin{dcases}
    \mathmakebox[10.5em][l]{(\minhatA{A_k+1},\xi-1)}
    \text{if $d = 1$;}\\[1ex]
    (\minhatA{A_k+1},\max\{R_k-1,0\})\\
    \qquad\qquad\text{ if $d=0$ and $R_k\neq 1$;}\\[1ex]
    (\minhatA{P+1},0)\\
    \qquad\qquad\text{ if $d=0$ and $R_k = 1$;}\\
    \end{dcases}
    \end{equation}
    \begin{equation}
        \begin{aligned}
        X_{k,f}=
            \begin{cases}
            (\minhatA{A_k+1},\xi)&\text{ if $d=1$;}\\
            (\minhatA{A_k+1},R_k)&\text{ if $d=0$.}
            \end{cases}
        \end{aligned}
    \end{equation}
    
    \item \textbf{Costs}: The immediate cost of the MDP under state $X(t)$, $C(X(t))$, is defined as the sum of AoI for all destinations. Thus, we have
    \begin{equation}
        C(X(t)) = \sum_{k\in\mathcal{K}} A_k(t).
    \end{equation}
\end{itemize}

Then, we define the stationary and deterministic policy.
\begin{definition}
A \textit{stationary} and \textit{deterministic} policy $\pi$ is a mapping from a state in the state space to a determined action. That is, $\pi: \mathcal{X} \to \mathcal{D}$. We define $\Pi$ as a set of stationary deterministic policies for the above MDP.\footnote{In general, a policy $\pi=\{\mu_1,\mu_2,...\}$ indicates the action for each time slot. Each $\mu_{t} \in \mathcal{M}$, usually referred to as a \textit{decision rule}, is a mapping from a state $X$ to an action $d$, and $\mathcal{M}$ is the set of {mappings} $\mu: \mathcal{X} \to \mathcal{D}$. Nevertheless, if we focus on \textit{stationary} policies {in the form} $\pi = \{\mu^\star, \mu^\star, ...\}$, the action will only depend on the current state, and the influence of time can be ignored. Thus, we define the policy as a mapping from a state to an action in our discussion.} 
\end{definition}

From \cite{Puterman}, a policy may take action based on all of its historical states. However, {finding such policies is challenging.} Therefore, we will investigate the optimal policy {within} the space of $\Pi$. For a policy $\pi\in\Pi$, the long-term time-average AoI is
\begin{equation}
    V^\pi(X_1)=\limsup_{T \to \infty} \frac{1}{T}\mathbbm{E}_{\pi}[\sum_{t=1}^T C(X(t))|X(1)=X_1],
\end{equation}
where $\mathbbm{E}_\pi$ denotes the expected value under the policy $\pi$, and $X_1$ is the system state vector of the initial state. For simplicity, we suggest that $X_1 = (\xi,(\xi,0)_{k\in\mathcal{K}})$ and eliminate $X_1$ subsequently.

\begin{definition}
    A policy $\pi^\star\in\Pi$ is an \textit{age-optimal policy} if it minimizes the long-term time-average AoI, as follows:
    \begin{equation}
        V^{\pi^\star} \triangleq \min_{\pi \in \Pi} V^\pi.
    \end{equation}
The \textit{optimal average AoI} is $V^\star \triangleq V^{\pi^\star}$.
\end{definition}
The objective of this paper is to find an \textit{age-optimal policy} $\pi^\star$ that minimizes the objective function $V^\pi$.

\section{Optimal policy and structural properties} \label{optimal policy}

{In this section, we first delve into how the unique behavior of AoI in our system allows us to eliminate certain states. Next, we explore how to find the optimal policy by solving the Bellman equation through the relative value iteration algorithm. However, facing the challenge of high complexity due to the curse of dimensionality, it becomes essential to identify structural properties that can simplify the algorithm. Finally, we integrate these insights to introduce a structure-aware relative value iteration algorithm (RVIA), designed to {efficiently} determine the optimal policy.}

\subsection{State space reduction}\label{Formal definition of the state space}
{Even though the state space size described above is large, some state vectors are unreasonable {when} we further explore the AoI transition model.} Therefore, we shall exclude them from the state space in our discussion. All system state vectors should satisfy the following three propositions.
\begin{proposition}\label{prop:Rk_ge_xi_minus_P_if_P_le_xi}
For any $ X \in \mathcal{X}$, if $\exists P \le \xi$, then $R_k \ge \xi - P,\forall k \in \mathcal{K}$. 
\end{proposition}

\begin{proof}
Intuitively, if we {transmitted only} $P \leq \xi$ packets previously, the {best-case scenario} for device $k$ is that all packets were successfully received, resulting in the minimum possible $R_k = \xi - P$. {If $R_k > \xi - P$, it} implies that one or more packets were lost.
\end{proof}

\begin{proposition}\label{prop:if_Rk_0_P_is_Ak_or_Ak_minus_P_ge_xi}
For any $ X \in \mathcal{X}$, if $\exists k \in \mathcal{K}$ such that $R_k=0$, then $P = A_{k}$. Otherwise, $A_k - P \ge \xi$ if $R_k>0$ and $A_k\neq\hat{A}$.
\end{proposition}

\begin{proof}
According to the age transition model, if $R_k=0$, it should have passed the transition with $X_{k}=(A_{k},1)$, action $d=0$ and $u_k = 1$ previously. From \eqref{AoI_model_d0_P} and \eqref{AoI_model_d0_Xk}, the new state after the transition is $P' = \minhatA{P+1}$ and $X_k' = (\minhatA{P+1},0)$, where $P' = A_k'$ and $R_k'=0$. {If the source has not yet generated a new status update, $R_k=0$ remains with $P$ and $A_k$ increasing by one simultaneously{;} thus, $P = A_k$ {is maintained}.}

{On the other hand, if the source decides to generate a new status update, {according to} \eqref{AoI_model_d1_P} and \eqref{AoI_model_d1_Ak}, the new state has $P' = 1$ and $A_k'=\min(\hat{A}, A_k+1)$.} Since the minimum of $A_k$ is $\xi$, we have $A_k'-P'\ge\xi$. Hereafter, $P$ and $A_k$ will grow together before $R_k$ becomes $0$, so $A_k'-P'\ge\xi$ holds before $A_k$ attains $\hat{A}$. If $A_k = \hat{A}$, $A_k$ will not grow, but $P$ may increase, so we cannot guarantee the inequality {holds} true.
\end{proof}

\begin{proposition}\label{prop:no_device_has_age_differing_more_than_xi}
For any $ X \in \mathcal{X}$, if $\exists k,k' \in \mathcal{K}$, such that $A_k \neq \hat{A}$, $A_{k'} \neq \hat{A}$, and $A_k \neq A_{k'}$, then $|A_k-A_{k'}| \ge \xi$.
\end{proposition}

\begin{proof}
Assume $d=1$ is taken on some states $X\in \mathcal{X}$. Then, we define the state in the next time slot as $\bar{X}=(1,(\bar{X_k})_{k\in\mathcal{K}})$, where the bar is denoted as the initial state here. From \eqref{AoI_model_d1_Ak}, $\bar{A_k} \geq \xi+1,\ \forall k\in\mathcal{K}$. {Note that the difference in $A_k$ between the two devices remains if {neither} of them successfully decodes the update.} Therefore, we assume device $k'$ has successfully received $\xi$ packets after $t-1$ transmissions of the same update, and we have the state $X=(t,(X_k)_{k\in\mathcal{K}})$ with $A_{k'}=t$. We now analyze the difference of AoI between device $k$ and $k'$, that is, the relationship of $A_k$ and $A_{k'}$. 

The first scenario is that device $k$ also successfully decoded the message in this time slot. Therefore, $A_k=A_{k'}=t$. 
The second scenario is device $k$ did not decode in this time slot, but $\hat{A} \ge t \ge \hat{A}-\xi$. In this case, since $\bar{A_k}\ge \xi+1$ initially, $A_k = \minhatA{\bar{A_k}+t-1} = \hat{A}$. Therefore, we have $\xi\ge A_k-A_{k'} \ge 0$. 
Finally, if $t<\hat{A}-\xi$ and the device $k$ did not decode the message in this time slot, we have $A_k = \bar{A_k}+t-1$. Hence, $A_k-A_{k'}=\bar{A_k}-1 \geq \xi$. The difference will remain until the next time of successful message decoding, or one of the device's age attains $\hat{A}$.

In conclusion, if $\exists k,k' \in \mathcal{K}$, such that $A_k \neq \hat{A}$, $A_{k'} \neq \hat{A}$, and $A_k \neq A_{k'}$, it implies the difference is caused by the third scenario, and thus $|A_k-A_{k'}| \ge \xi$.
\end{proof}

\begin{definition}
A \textit{system state vector space} $\mathcal{X}$ is composed of all possible combinations of $(P,(A_k,R_k)_{k
\in\mathcal{K}})\in\mathcal{P} \times \prod_{k\in\mathcal{K}}(\mathcal{A} \times \mathcal{R})$, and the combinations should satisfy the three propositions.
\end{definition}

\subsection{Bellman equation}
After clearly defining the MDP formulation of this problem, we can obtain an age-optimal policy by solving the Bellman equation.
\begin{lemma}\label{lem: Existance of Bellman equation}
There exists a scalar $\lambda$ and a function $h(X)$ that satisfy the following Bellman equation: 
\begin{equation}\label{Bellman equation}
    \lambda+h(X)= \sum_{k\in\mathcal{K}}A_k+\min_{d\in\mathcal{D}} \sum_{X'\in\mathcal{X}}Pr[X'|X,d]h(X'),
\end{equation}
where $\lambda$ is the value of optimal average AoI. $\lambda$ is independent of the initial state and can be written formally as 
\begin{equation}
    \lambda = \lim_{\gamma \to 1} (1-\gamma)V_{\gamma}(X), \forall X \in\mathcal{X},
\end{equation}
and $h(X)$ is the relative value function for any given fixed state $o$, satisfies
\begin{equation}
    h(X) = \lim_{\gamma \to 1}(V_{\gamma}(X)-V_{\gamma}(o)), \forall X\in\mathcal{X},
\end{equation}
where $V_{\gamma}(X)$ is the value of the state $X$ under the discount factor $\gamma$ and the optimal policy $\pi$, given as
\begin{equation}\label{discounted optimal value}
    V_{\gamma}(X) = \min_{\pi\in\Pi}\limsup_{T \to \infty} \frac{1}{T}E_{\pi}[\sum_{t=1}^T \gamma^{t-1}C(X(t))|X(1)=X].
\end{equation}

In addition, if a policy $\pi^\star(X)$ attains the minimum of the right hand side in equation \eqref{Bellman equation} for all $X$, that is, 
\begin{equation}\label{definition of the optimal action}
    \pi^\star(X)=\arg\min_{d\in\mathcal{D}} \sum_{X'\in\mathcal{X}}Pr[X'|X,d]h(X'), \forall X\in\mathcal{X},
\end{equation}
then the stationary policy $\pi^\star$ is age-optimal.
\end{lemma}
\begin{proof}
In our problem, we have to prove the MDP is \textit{communicating}. That is, for every two states $X,X' \in \mathcal{X}$, there exists a stationary policy $f$ such that for some $t$, we have $Pr[X(t)=X'|X(0)=X,f]>0$. By \cite[Definition 4.2.2, Propositions 4.2.1 and 4.2.3]{Bertsekas}, if a MDP is communicating, the weak accessibility condition holds for the MDP. Then, we can prove the optimal average AoI is the same for all initial states, so \eqref{Bellman equation} to \eqref{discounted optimal value} hold. Besides, \cite[Corollary 8.2.5]{Puterman} uses Laurent series expansion of $V_\gamma$ to prove $\lambda = V^\star(X_1), \forall X_1\in\mathcal{X}$. We can also find an age-optimal policy from the Bellman equation in \eqref{definition of the optimal action}, according to \cite[Theorem 8.4.3]{Puterman}. We put our proof about the communicating property of MDP in Appendix~I. 
\end{proof}

From \eqref{definition of the optimal action}, the optimal policy depends on the relative value function $h(X)$. However, there is generally no closed-form solution for $h(X)$. Alternatively, we tend to perform the relative value iteration algorithm (RVIA) \cite[Section 3.4.2]{powell2007approximate} to solve $h(X)$ numerically. By initializing $h_0(X)=0$ for all $X \in \mathcal{X}$ and starting with any fixed state $o \in \mathcal{X}$, a single iteration in RVIA {is as follows}.
\begin{equation}\label{Q-factor-iter}
    Q_{n+1}(X,d)=\sum_{k\in\mathcal{K}}A_k+\sum_{X'\in\mathcal{X}}Pr[X'|X,d]h_n(X') - h_n(o)\text{;}
\end{equation}

\begin{equation}
\begin{aligned}\label{relative-function-iter}
h_{n+1}(X) & = \min_{d\in\mathcal{D}}Q_{n+1}(X,d),
\end{aligned}
\end{equation}
where $Q_n(X,d)$ and $h_n(X)$ are Q-factor and relative value function at iteration $n$, respectively. As described in \cite[Section 9.5.3]{Puterman}, $h_{n+1}(X)-h_n(X)$ will converge to 0 if the MDP is weakly communicating\footnote{From \cite{Puterman,bedewy2021optimalSampling}, the sufficient condition for the convergence of RVIA is the aperiodicity of the Markov chain induced by the age-optimal policy $\pi^\star$. However, we can apply the aperiodicity transform mentioned in \cite[Section 8.5.4]{Puterman} to ensure the RVIA is available. Therefore, the proof of aperiodicity of the Markov chain is not necessary here.}. That is,
\begin{equation}\label{Take limit to h_n}
    \lim_{n\to\infty}h_n(X)=h(X), \forall X\in\mathcal{X}.
\end{equation}
The Equation \eqref{Take limit to h_n} also implies the convergence of $Q_n(X,d)$. Hence, we have $\lim_{n\to\infty}Q_n(X,d)=Q(X,d), \forall X\in\mathcal{X},d\in\mathcal{D}$. Then, we can rewrite \eqref{Bellman equation} as
\begin{equation}\label{Bellman equation for Q-value}
   \lambda + Q(X,d)= \sum_{k\in\mathcal{K}}A_k+ \sum_{X'\in\mathcal{X}}Pr[X'|X,d]h(X'),    
\end{equation}
where $Q(X,d)$ can be called as Q-factor. Furthermore, the optimal average AoI is the relative value function of the fixed state, $\lambda = h(o)$, by comparing \eqref{Q-factor-iter} and \eqref{Bellman equation for Q-value}.

\subsection{Structural Properties}
In this section, we will further investigate the special structural properties in the MDP $\Delta$ to reduce the complexity of the RVIA. We define $\mathbf{A} = (A_k)_{k\in\mathcal{K}}$ and $\mathbf{R} = (R_k)_{k\in\mathcal{K}}$ as the {vectors} of $A_k$ and $R_k$ for all $k\in\mathcal{K}$. The symbol $\preceq$ denotes vector inequality in $\mathbb{R}^K$: $\mathbf{A}^1 \preceq \mathbf{A}^2$ means $A_k^1 \leq A_k^2$ for all $k\in\mathcal{K}$. Then, we start {with} the property of the relative value function.
\begin{lemma}\label{lem:increasing_relative_value_function}
For any $X^1 = (P^1,\mathbf{A}^1,\mathbf{R}^1), X^2 = (P^2,\mathbf{A}^2,\mathbf{R}^2) \in \mathcal{X}$, if they satisfy $P^1 \le P^2$, $\mathbf{A}^1 \preceq \mathbf{A}^2$ and $\mathbf{R}^1\preceq\mathbf{R}^2 $, then $h_n(X^1)\le h_n(X^2)$ for all $n$. Hence, $h(X^1)\le h(X^2)$.
\end{lemma}
\begin{proof}
See Appendix B.
\end{proof}
\cref{lem:increasing_relative_value_function} indicates the relative value function of the state is monotonically increasing in $P$, $\mathbf{A}$ and $\mathbf{R}$. This implies state $(\hat{A},(\hat{A},\xi)_{k\in\mathcal{K}})$ is the worst state among all possible states. The lemma is essential to prove \cref{thm:structure_prop_of_Q} and the structural properties of the optimal policy, as we can see in the following proof.

\begin{theorem}\label{thm:structure_prop_of_Q}
For any $X^1 = (P^1,\mathbf{A}^1,\mathbf{R}^1), X^2 = (P^2,\mathbf{A}^2,\mathbf{R}^2) \in \mathcal{X}$, if they satisfy $\mathbf{A}^1 = \mathbf{A}^2$, then $Q_n(X^1,1) = Q_n(X^2,1)$ for all $n$. Therefore, $Q(X^1,1) = Q(X^2,1)$. Moreover, if $P^1 \le P^2$ and $\mathbf{R}^1 \preceq \mathbf{R}^2$, $Q_n(X^1,0) \le Q_n(X^2,0)$ for all $n$. Hence, $Q(X^1,0) \le Q(X^2,0)$.
\end{theorem}
\begin{proof}
See Appendix C.
\end{proof}

When calculating the Q-factor, applying this theorem can reduce the complexity of RVIA. Originally, one has to calculate $Q_n(X,1)$ for all $X\in\mathcal{X}$ in each iteration. However, {according to} \cref{thm:structure_prop_of_Q}, states with the same vector $\mathbf{A}$ will have the same $Q_n(X,1)$ regardless of $P$ and $\mathbf{R}$. Therefore, we can first find all possible values of $Q_n(X,1)$ according to different $\mathbf{A}$ and store the results in a table indexed by $\mathbf{A}$. Then, when we need to update $Q_n(X,1)$, {we can do so from the table}. This property will be exploited in \cref{low-complexity suboptimal algorithm}. By combining \cref{lem:increasing_relative_value_function} and \cref{thm:structure_prop_of_Q}, we can further obtain the structural property of the optimal policy.
\begin{theorem}\label{thm:structure_prop_with_same_A}
For any $X^1 = (P^1,\mathbf{A}^1,\mathbf{R}^1), X^2 = (P^2,\mathbf{A}^2,\mathbf{R}^2) \in \mathcal{X}$ with $P^1 \le P^2$, $\mathbf{A}^1 = \mathbf{A}^2$ and $\mathbf{R}^1\preceq\mathbf{R}^2 $, if $\pi^\star(X^2) = 0$, then $\pi^\star(X^1) = 0$. Similarly, if $\pi^\star(X^1) = 1$, then $\pi^\star(X^2) = 1$.
\end{theorem}
\begin{proof}
From \cref{thm:structure_prop_of_Q}, we have $Q(X^1,1) = Q(X^2,1)$ and $Q(X^1,0) \le Q(X^2,0)$. This follows the inequality 
\begin{equation}\label{structure_Q}
    Q(X^1,0) - Q(X^1,1) \le Q(X^2,0) - Q(X^2,1)
\end{equation}
In addition, $\pi^\star(X^2) = 0$ implies $Q(X^2,0) - Q(X^2,1) \le 0$. Combining the above inequality, we obtain $Q(X^1,0) - Q(X^1,1) \le 0$, which is equal to $\pi^\star(X^1) = 0$. 

For the case that if $\pi^\star(X^1) = 1$, then $\pi^\star(X^2) = 1$, we can multiply $-1$ to \eqref{structure_Q} and applying $Q(X^1,1) - Q(X^1,0) \le 0$. Hereafter, the proof is completed.
\end{proof}

\cref{thm:structure_prop_with_same_A} reveals the unique structure of this system and enables us to improve efficiency when executing RVIA. If we can find the optimal action for a state, then many optimal actions for other states can be derived from this theorem. Moreover, \cref{prop:R_is_xi_or_0_then_pi_is_1} indicates another structural property for an age-optimal policy.

\begin{proposition}\label{prop:R_is_xi_or_0_then_pi_is_1}
If a state $X\in\mathcal{X}$ has $\mathbf{R}=\pmb{\xi}$ or $\mathbf{R}=\textbf{0}$, $\pi^\star(X)=1$.
\end{proposition}
\begin{proof}
See Appendix D.
\end{proof}
In addition to the rigorous proof, we can intuitively understand the above proposition. If all devices fail to receive a packet or {successfully decode an update}, the source should start {transmitting} a new status update. We can use this property to simplify the calculation of {the} optimal policy for some states.

\subsection{Solution of the optimal policy with structural properties}
Theoretically, we can obtain $h(X)$ for all $X$ by RVIA and thus obtain the optimal policy. However, {performing the algorithm will take a considerable amount of time} due to the curse of dimensionality \cite{powell2007approximate}. As a result, we begin by applying the structural properties to reduce its complexity.
\begin{algorithm}
  \caption{Structure-aware RVIA}\label{optimal_RVIA_algorithm}
  \SetKwRepeat{Do}{do}{while}%
  \SetKwRepeat{elseIf}{If}{else}%

  \textbf{Initialize}: $h_0(X) = 0, \forall X\in\mathcal{X}$, a fixed state $o\in\mathcal{X}$,
  $n=0$ and a tolerance $\epsilon>0$
  
  \Do{$\max_{X\in\mathcal{X}}|h_{n}(X)-h_{n-1}(X)|>\epsilon$}
  {
    
     \For{\textbf{each} $X=(P,\mathbf{A},\mathbf{R})\in\mathcal{X}$}
        {
        \uIf{$\mathbf{R} = \textbf{0}$ or $\mathbf{R} = \pmb{\xi}$}
        {
        $\pi^\star_n(X) \xleftarrow[]{} 1$
        }
        \uElseIf{there exists a non-negative integer $c$ and K-dimensional non-negative integers $\textbf{c}$ such that $\pi^\star(P+c,\mathbf{A},\mathbf{R}+\textbf{c}) = 0$  }
        {
            $\pi^\star_n(X) \xleftarrow[]{} 0$
        }
        \uElseIf{there exists a non-negative integer $c$ and K-dimensional non-negative integers $\textbf{c}$ such that $\pi^\star_n(P-c,\mathbf{A},\mathbf{R}-\textbf{c}) = 1$ }
        {
            $\pi^\star_n(X) \xleftarrow[]{} 1$
        }
        \Else
        {
            $\pi^\star_n(X) \xleftarrow[]{} \arg\min_{d\in\mathcal{D}}
            \sum_{X'\in\mathcal{X}}Pr[X'|X,d]h_n(X')$
        }
        $h_{n+1}(X) \xleftarrow[]{} \sum_{k\in\mathcal{K}}A_k\break
            +\sum_{X'\in\mathcal{X}}Pr[X'|X,\pi^\star_n(X)]h_n(X') - h_n(o)$

        }
    $n \xleftarrow[]{} n+1$
}
$\pi^\star \xleftarrow[]{} \pi^\star_n$
\end{algorithm}
The structure-aware RVIA is proposed in \cref{optimal_RVIA_algorithm}. The algorithm takes advantage of the \cref{prop:R_is_xi_or_0_then_pi_is_1} and \cref{thm:structure_prop_with_same_A}. In \cref{optimal_RVIA_algorithm}, we initialize the relative value function for all states to zero and select a fixed state $o$ and tolerance $\epsilon$. Hereafter, in each iteration, we seek the optimal action $\pi^\star(X)$ for each state $X\in\mathcal{X}$. If the state has the vector $\mathbf{R} = \mathbf{0}$ or $\mathbf{R} = \pmb{\xi}$, the optimal action can be directly determined by \cref{prop:R_is_xi_or_0_then_pi_is_1}. Otherwise, we will check whether we can apply the structural properties in \cref{thm:structure_prop_of_Q}. {We can identify its optimal action if the structural properties are available, as in Lines 7 and 9.} Then, if we can not apply all structural properties, we will calculate it directly in Line 11. After determining the optimal action for state $X$, we can update the relative value function of state $X$ in Line 12. The algorithm will continue until the maximum difference between the value of the relative value function in the current and that in the last iteration is less than $\epsilon$. From \eqref{Take limit to h_n}, this algorithm is guaranteed to converge.

Upon calculating the optimal action for all possible states before transmission begins, the source can determine its action at time slot $t$ based on the state $X(t)$. Nevertheless, since we have to iterate through all possible states, the complexity is still high at $\mathcal{O}(\prod_{k\in\mathcal{K}}\hat{A}(\hat{A}-\xi+1)(\xi+1))$. The algorithm may not be feasible as the number of devices rises because the time complexity increases exponentially. Hence, we investigate an approximation method in the next section to overcome this issue.

\section{Low Complexity suboptimal solution} \label{Section low-complexity}
\subsection{Decomposition of Bellman equation}
To resolve the curse of dimensionality, we will approximate $Q(X,d)$. Before the approximation, we first define $\widetilde{X}_k \triangleq (P,X_k)$ as the \textit{extended state vector of device} $k$. $\widetilde{X}_k \in \widetilde{\mathcal{X}}_k \subset \mathcal{P} \times \mathcal{X}_k$ is the extended state space of device $k$. The concept behind this definition is that a device $k$ should know the age of the status update $P$ to update its local age $A_k$. However, only the source knows the system state of all devices. Therefore, we can {consider} the extended state as local information for a single device. Based on the extended state of device $k$, we formulate a new MDP {denoted} as $\Delta_k$ for each device $k$. It can be regarded as the original MDP $\Delta$ with $K=1$ and packet erasure rate $p_k$, comprising the following components:
\begin{itemize}
    \item \textbf{States}: At the beginning of each time slot $t$, the extended state of device $k$ can be described as
    \begin{equation}
    \begin{aligned}
        \widetilde{X}_k(t) &= (P(t),X_k(t)) = (P(t),A_k(t),R_k(t)),  
    \end{aligned}
    \end{equation}
    
    \item \textbf{actions}: At each time slot $t$, the source will select a decision $d(t)\in\mathcal{D}=\{0,1\}$.
    
    \item \textbf{Transition probabilities}: Since there is only one device, the transition probability can be simply written as
    \begin{equation}\label{Transition prob in extended space}
        \begin{aligned}
        Pr[\widetilde{X}_k'|\widetilde{X}_k,d]=
            \begin{cases}
            1-p_k&\widetilde{X}_k'=\widetilde{X}_{k,s}\text{;}\\
            p_k&\widetilde{X}_k'=\widetilde{X}_{k,f}\text{;}\\
            0&\text{otherwise.}
            \end{cases}
        \end{aligned}
    \end{equation}
    
    Because the transition of the packet's age is deterministic, the transition probability will not be affected by $P$. Hence, \eqref{Transition prob in extended space} essentially has the same structure as in \eqref{Transition prob in ori}. We only need to slightly modify the states indicating successful and failed reception as follows:
    
    \begin{equation}
    \widetilde{X}_{k,s}=
    \begin{dcases}
    \mathmakebox[10.5em][l]{(1,\minhatA{A_k+1},\xi-1)}
    \text{if $d = 1$;}\\[1ex]
    (\minhatA{P+1},\minhatA{A_k+1},\max\{R_k-1,0\})\\
    \qquad\qquad\text{ if $d=0$ and $R_k\neq 1$;}\\[1ex]
    (\minhatA{P+1},\minhatA{P+1},0)\\
    \qquad\qquad\text{ if $d=0$ and $R_k = 1$;}\\
    \end{dcases}
    \end{equation}
    \begin{equation}
        \begin{aligned}
        \widetilde{X}_{k,f}=
            \begin{cases}
            (1,\minhatA{A_k+1},\xi)&\text{ if $d=1$;}\\
            (\minhatA{P+1},\minhatA{A_k+1},R_k)&\text{ if $d=0$.}
            \end{cases}
        \end{aligned}
    \end{equation}
    \item \textbf{Costs}: The immediate cost of the new MDP under state $\widetilde{X}_k(t)$ is simply the current AoI of the device.
    \begin{equation}
        C(\widetilde{X}_k(t)) = A_k(t)
    \end{equation}
\end{itemize}

\begin{remark}\label{rem:inherit_main_MDP}
The new MDP $\Delta_{k}$ can be viewed as the original MDP $\Delta$ with only one device ($K=1$) in the network. The device's packet erasure rate equals $p_k$. Therefore, the above propositions, lemmas, and theorems in $\Delta$ also hold for $\Delta_k$.
\end{remark}

Notice that our target is not to find the optimal policy for $\Delta_k$. However, the new MDP $\Delta_k$ can take advantage of the properties we obtained in the original MDP $\Delta$. Now, we approximate $Q(X,d)$ as 
\begin{equation} 
    Q(X,d) \approx \sum_{k\in\mathcal{K}} Q_k(\widetilde{X}_k,d),
\end{equation}
where $Q_k(\widetilde{X}_k,d)$ satisfies
\begin{equation}\label{Bellman equation for Q_k}
\begin{aligned}
        \lambda_k &+ Q_k(\widetilde{X}_k,d)= \\
    &A_k+ \sum_{\widetilde{X}_k'\in\widetilde{\mathcal{X}}_k}Pr[\widetilde{X}_k'|\widetilde{X}_k,d]\min_{d'\in\mathcal{D}}Q(\widetilde{X}_k',d'),\ \forall \widetilde{X}_k\in\widetilde{\mathcal{X}}_k.
\end{aligned}
\end{equation}

We can guarantee Equation \eqref{Bellman equation for Q_k} holds by applying \cref{lem: Existance of Bellman equation} and \cref{rem:inherit_main_MDP}. Hereafter, we propose a suboptimal policy under the approximation as
\begin{equation} \label{suboptimal}
\widetilde{\pi}^\star(X)=\arg\min_{d\in\mathcal{D}} \sum_{k\in\mathcal{K}} Q_k(\widetilde{X}_k,d).
\end{equation}

The suboptimal policy now depends on $Q_k(\cdot,\cdot)$. Because \eqref{Bellman equation for Q_k} is essentially a Bellman equation, we can also solve $Q_k(\cdot,\cdot)$ by RVIA like an analogy of \eqref{Q-factor-iter} and \eqref{relative-function-iter}, where 
\begin{equation}\label{Q-factor-iter-single}
\begin{aligned}
        Q_k^{n+1}(&\widetilde{X}_k,d) = \\
    &A_k+ \sum_{\widetilde{X}_k'\in\widetilde{\mathcal{X}}_k}Pr[\widetilde{X}_k'|\widetilde{X}_k,d]h_k^n(\widetilde{X}_k') - h_k^n(o),
\end{aligned}
\end{equation}
\begin{equation}\label{relative-function-iter-single}
h_k^{n+1}(\widetilde{X}_k) = \min Q_k^{n+1}(\widetilde{X}_k)
\end{equation}
where $Q_k^{n}(X,d)$ and $h_k^{n}(X)$ are Q-factor and relative value function at iteration $n$ for $\Delta_k$, and $o\in\widetilde{\mathcal{X}}_k$ is prescribed. We can guarantee the convergence of each $Q_k(\widetilde{X}_{k},d)$ based on  \eqref{Take limit to h_n}. By repeating this process for different $p_k$, we can obtain the suboptimal policy. The structural properties of the optimal policy are not available here. This is because the optimal decision for $k$ might not be optimal for the entire system. Both the Q-factors for decisions 0 and 1 are required for the source to determine the suboptimal policy, as indicated by \eqref{suboptimal}. That is, lines 4 to 11 will be ignored, and line 12 will be replaced by calculating the Q-factor for both decisions 0 and 1 in \cref{optimal_RVIA_algorithm}. Interestingly, the suboptimal policy also has the same structure properties as \cref{thm:structure_prop_with_same_A}:
\begin{corollary}\label{cor:structural_prop_for_small_MDP}
For any $X^1 = (P^1,\mathbf{A}^1,\mathbf{R}^1), X^2 = (P^2,\mathbf{A}^2,\mathbf{R}^2) \in \mathcal{X}$ with $P^1 \le P^2$, $\mathbf{A}^1 = \mathbf{A}^2$ and $\mathbf{R}^1\preceq\mathbf{R}^2 $, if $\widetilde{\pi}^\star(X^2) = 0$, then $\widetilde{\pi}^\star(X^1) = 0$. Similarly, if $\widetilde{\pi}^\star(X^1) = 1$, then $\widetilde{\pi}^\star(X^2) = 1$.
\end{corollary}
\begin{proof}
For every $\widetilde{X}_k^1,\widetilde{X}_k^2$ derived from $X^1,X^2$, they also have $P^1 $ $\le$ $P^2$, $A_k^1=A_k^2$ and $R_k^1\le R_k^2$. Hence, from \cref{thm:structure_prop_of_Q}, $Q_k(\widetilde{X}_k^1,0) \le Q_k(\widetilde{X}_k^2,0)$ and $Q_k(\widetilde{X}_k^1,1) = Q_k(\widetilde{X}_k^2,1)$. Since the result holds for all devices $k\in\mathcal{K}$, the inequality remains when we add them up. Furthermore, if $\widetilde{\pi}^\star(X^2) = 0$, it implies $\sum_{k\in\mathcal{K}} Q_k(\widetilde{X}_k^2,1)$ $\ge$ $\sum_{k\in\mathcal{K}} Q_k(\widetilde{X}_k^2,0)$. By $\sum_{k\in\mathcal{K}} Q_k(\widetilde{X}_k^1,1)$ $=$ $\sum_{k\in\mathcal{K}} Q_k(\widetilde{X}_k^2,1)$ $\ge$ $\sum_{k\in\mathcal{K}} Q_k(\widetilde{X}_k^2,0) \ge \sum_{k\in\mathcal{K}} Q_k(\widetilde{X}_k^1,0)$, we thus get $\widetilde{\pi}^\star(X^1) = 0$. The structural property of suboptimal policy for action $1$ can be proven in a similar way.
\end{proof}

We will further explain how to apply the structural properties for calculating the Q-factor and the suboptimal policy to reduce the algorithm's complexity in the next subsection.

\subsection{Decentralized structure-aware low-complexity RVIA}\label{low-complexity suboptimal algorithm}
Based on \eqref{Q-factor-iter-single}, \eqref{relative-function-iter-single} and \cref{thm:structure_prop_of_Q}, we propose a decentralized structure-aware low-complexity RVIA for the suboptimal policy, as shown in \cref{lowcomp_RVIA_algorithm}. This algorithm can be separated into two phases. From \eqref{suboptimal}, we know the suboptimal policy can be obtained by calculating the Q-factors under different separated extended states of each device $k$. Therefore, in the first phase, we obtain the Q-factors of each $\Delta_k$. Since we aim to get the Q-factors instead of the optimal policy under $\Delta_k$, \cref{thm:structure_prop_with_same_A} cannot be applied. However, we can utilize \cref{thm:structure_prop_of_Q}. We have proven that for all states with the same age vectors, the Q-factors for action 1 are the same. Therefore, we can first define some reference extended states with different age vectors\footnote{In practice, we can set a reference extended state $\widetilde{X}_{k,ref} = ((\xi-1)\mathbbm{1}(A_k=\xi)+1,A_k,\xi(1-\mathbbm{1}(A_k=\xi)))$, {because the only possible extended state is $(\xi,\xi,0)$ if $A_k=\xi$.} Otherwise, we can select the reference extended state as $(1,A_k,\xi)$. Based on Propositions 1 to 3, we can verify the arrays are in the extended state space.} and find their Q-factors, as we can see in Line 3 to Line 5. The calculated results are stored in a table $f_k^n[\cdot]$, and will be assigned to the corresponding state in Line 8. By doing so, we only have to perform $\hat{A}-\xi+1$ times of calculation to obtain $Q_k^n(\widetilde{X}_k,1)$ for all $\widetilde{X}_k\in\widetilde{\mathcal{X}}_k$.

The complexity of the low-complexity algorithm for the source reduces from $\mathcal{O}(\prod_{k\in\mathcal{K}}\hat{A}(\hat{A}-\xi+1)(\xi+1))$ to $\mathcal{O}(\sum_{k\in\mathcal{K}}\hat{A}(\hat{A}-\xi+1)(\xi+1))$. Moreover, as we mentioned that $\widetilde{X}_{k}$ can be considered local information for device $k$, we find that the computation is not restricted to perform on the source. Instead, it is possible to implement this computation on the device side, especially since many devices have the ability to perform fundamental calculations nowadays. This decentralized feature, which allows parallel computing, further reduces execution time. If the computation can be finished on each device, the source does not need to perform any calculation in this phase, and the complexity for every single device is $\mathcal{O}(\hat{A}(\hat{A}-\xi+1)(\xi+1))$.

If we adopt the decentralized algorithm, the Q-factors are stored in each device. Therefore, devices shall report the results of Q-factors to the source. Then, the source computes the suboptimal policy based on the Q-factors in phase 2, Line 21. Since the structural property of {the} suboptimal policy exists, the source does not need to add Q-factors for all states to get the optimal action, as shown in Line 14 to Line 19.

{\begin{remark}
    Although the original RVIA has high complexity, we emphasize the value of studying the optimal policy. The algorithm serves as a theoretical benchmark, providing insights into the performance limits of low-complexity methods. Moreover, it aids in designing low-complexity RVIA by providing theoretical support. Additionally, the optimal policy reveals key system dynamics, such as the impact of packet erasure rates on network performance, as we will demonstrate in \cref{opt_subopt_comparison}.
\end{remark}}

\begin{algorithm}

  \caption{Decentralized structure-aware low-complexity RVIA}\label{lowcomp_RVIA_algorithm}
  
  \SetKwRepeat{Do}{do}{while}%

  \nonl\textbf{PHASE 1: RVIA at each device $k$ based on $\Delta_k$}: Before the source starts the transmission, each device $k$ calculates its Q-factors $Q_k(\widetilde{X}_k,d)$ separately:

  \textbf{Initialize}: $h_k^0(\widetilde{X}_k) = 0, \forall \widetilde{X}_k \in\widetilde{\mathcal{X}}_k$, a fixed state $o\in\widetilde{\mathcal{X}}_k$,
  $n=0$ and tolerance $\epsilon>0$
  
  \Do{$\max_{\widetilde{X}_k\in\widetilde{\mathcal{X}}_k}|h_k^{n}(\widetilde{X}_k)-h_k^{n-1}(\widetilde{X}_k)|>\epsilon$}
  {
     \For{\textbf{each} $A_k\in\mathcal{A}$}
        {
        Select a reference extended state $\widetilde{X}_{k,ref} = (P_0,A_k,R_{k,0})\in\widetilde{\mathcal{X}_k}$  \;
        $f_k^{n+1}[A_k] \xleftarrow[]{} A_k +\break \mathbbm{E}[h_k^{n}(\widetilde{X}_k')|\widetilde{X}_{k,ref},d=1] - h_k^n(o)$\;
        }
    \For{\textbf{each} $\widetilde{X}_k\in\widetilde{\mathcal{X}}_k$}
    {
        $Q_k^{n+1}(\widetilde{X}_k,0) \xleftarrow[]{} A_k + \break \mathbbm{E}[h_k^{n}(\widetilde{X}_k')|\widetilde{X}_k,d=0] - h_k^n(o)$\;
        $Q_k^{n+1}(\widetilde{X}_k,1) \xleftarrow[]{} f_k^{n+1}[A_k]$\;
        $h_k^{n+1}(\widetilde{X}_k) \xleftarrow[]{} \min_{d\in\mathcal{D}}Q_k^{n+1}(\widetilde{X}_k,d)$\;
    }
        
    $n \xleftarrow[]{} n+1$
}
$Q_k(\widetilde{X}_k,d) \xleftarrow[]{} Q_k^n(\widetilde{X}_k,d)$  $\forall \widetilde{X}_k\in\widetilde{\mathcal{X}_k}, d\in\mathcal{D}$

  \nonl\textbf{PHASE 2: The source obtains the suboptimal policy}: The source acquires the suboptimal policy based on reported Q-factors $Q_k(\widetilde{X}_k,d)$ from devices:
  
  \For{\textbf{each} $X\in\mathcal{X}$}{
  \uIf{$\mathbf{R}=\textbf{0}$ or $\mathbf{R}=\pmb{\xi}$}
  {
    $\widetilde{\pi}^\star(X) \xleftarrow[]{} 1$\;  }
  \uElseIf{there exists a non-negative integer $c$ and K-dimensional non-negative integers $\textbf{c}$ such that $\widetilde{\pi}^\star(P+c,\mathbf{A},\mathbf{R}+\textbf{c}) = 0$  }
        {
            $\widetilde{\pi}^\star(X) \xleftarrow[]{} 0$
        }
        \uElseIf{there exists a non-negative integer $c$ and K-dimensional non-negative integers $\textbf{c}$ such that $\widetilde{\pi}^\star(P-c,\mathbf{A},\mathbf{R}-\textbf{c}) = 1$ }
        {
            $\widetilde{\pi}^\star(X) \xleftarrow[]{} 1$
        }
  \Else{
  $\widetilde{\pi}^\star(X) \xleftarrow[]{} \arg\min_{d\in\mathcal{D}} \sum_{k\in\mathcal{K}} Q_k(\widetilde{X}_k,d)$
  }
  }
\end{algorithm}

\section{Performance Evaluation} \label{Section simulation}
    In this section, we shall compare our proposed optimal and low-complexity suboptimal solution with other reference baselines. First, we investigate the difference between the optimal and the proposed low-complexity suboptimal policies under four states and system settings. Afterward, we simulate the performance of AoI under different policies concerning packet erasure rates and the number of devices.

    \subsection{Performance Evaluation Methodologies}
    \begin{algorithm}
          \caption{Numberical method for searching optimal FR strategy threshold $P^\star$}\label{optimal_FR_strategy_threshold_algorithm}
          \SetKwRepeat{Do}{do}{while}%
          \SetKwRepeat{elseIf}{If}{else}%

        \KwResult{The optimal FR strategy threshold $P^\star$ such that the average AoI can be minimized.}
        \(\text{minAoI} \leftarrow \infty\) \;
        \(P^\star \leftarrow 0\) \;
        \For{\(n \leftarrow \xi\) \KwTo \(\hat{A}\)}{
            \( \text{sumAoI} \leftarrow 0 \) \;
            \For{\(k \in \mathcal{K}\)}{
                \( \text{sumAoI} \leftarrow \text{sumAoI} + V^{\pi_{\text{FR}}}(n, p_k) \) \;
            }
            \(\text{currAoI} \leftarrow \frac{\text{sumAoI}}{K}\) \;
            \If{\(\text{currAoI} < \text{minAoI}\)}{
                \(\text{minAoI} \leftarrow \text{currAoI}\) \;
                \(P^\star \leftarrow n\) \;
            }
        }
        \Return \(P^\star\) \;
        \end{algorithm}

    The simulations are conducted with MATLAB. For \cref{opt_subopt_comparison}, we perform \cref{optimal_RVIA_algorithm,lowcomp_RVIA_algorithm} to obtain the actions under different states. Since the Figs.~\ref{fig:Structure_prop}(a) and~\ref{fig:Structure_prop}(b), and Figs.~\ref{fig:Structure_prop}(c) and~\ref{fig:Structure_prop}(d) have different packet erasure rates, the algorithms are performed twice.

   In \cref{AoI_result}, we compare the AoI performance with respect to different combinations of packet erasure rates and different numbers of users. All of the simulations in the section have the system parameters $\hat{A} = 40$, $\xi = 10$ and $T = 50000$. Other parameter settings are listed in \cref{tab:parameters_for_6_3}. The results are averaged over 100 runs. Except for the optimal and low-complexity suboptimal solutions, we further compare their performance with three available baseline policies. 
   
   The {Optimal Fixed Redundancy (Optimal FR)} strategy does not require device feedback. The system calculates how many packets it should transmit based on the {static} packet erasure rates. If the number of currently transmitted packets is larger {or equal to} the threshold, it will start to send a new update in the next time slot.
   {In the study introducing the FR strategy \cite{yates2017timely}, the researchers detailed the average AoI for a specific device experiencing a packet erasure rate of $\delta$ when employing the FR strategy. This is represented by the equation:
   \begin{equation}
        V^{\pi_{\text{FR}}}(n, \delta) = \frac{n}{F_\xi(n)} - \frac{n}{2} + \frac{\xi F_{\xi+1}(n+1)}{(1-\delta)F_\xi(n)},
    \end{equation}
   where $n$ is the FR strategy threshold, $\xi$ is the rateless threshold, $F_\xi(n)$ is the Cumulative Distribution Function (CDF) of a negative binomial $(\xi, 1-\delta)$ random variable. The optimal FR strategy will find the optimal FR strategy threshold, denoted by $P^\star$, to minimize the average AoI for all devices. Since the possible threshold values range from $\xi$ to $\hat{A}$, and the status update for all users is independent and does not influence one another, the optimal FR strategy threshold $P^\star$ can be searched using {a numerical method as described} in \cref{optimal_FR_strategy_threshold_algorithm}.}
   
   On the other hand, the infinite incremental redundancy (IIR) strategy will not transmit a new update until all devices successfully receive more or equal to $\xi$ packets. This strategy can also be regarded as the persistent policy in \cite{Feng2019Ageoptimal}. The {Optimal Fixed Incremental Redundancy strategy (optimal FIR)} combines the optimal FR strategy and the IIR strategy. In summary, we can write the three strategies precisely as
    \begin{equation}
        \pi_{\text{FR}}(X) = 
        \begin{cases}
        1 & P \ge P^\star\\
        0 & P < P^\star
        \end{cases}
    \end{equation}
    \begin{equation}
                \pi_{\text{IIR}}(X) = 
        \begin{cases}
        1 &\ \mathbf{R} = \textbf{0}\\
        0 &\text{ otherwise}
        \end{cases}
    \end{equation}
    \begin{equation}
            \pi_{\text{FIR}}(X) = 
    \begin{cases}
    1 &\ \mathbf{R} = \textbf{0}\ \text{or}\ P \ge P^\star\\
    0 &\text{ otherwise}
    \end{cases}
    \end{equation}
    
        \begin{table*}
        \centering
        \caption{Parameter settings for \cref{AoI_result}.}
        
        \begin{threeparttable}
    \begin{tabular}{@{}l|c|c|c|c|c|c@{}}
    \toprule
    Figure    & \ref{fig:AoI_err_prob_1}  & \ref{fig:AoI_err_prob_2}  & \ref{fig:AoI_err_prob_3} & \ref{fig:same_err} & \ref{fig:two_err}  & \ref{fig:linear_err}  \\ \midrule
    Number of devices, $K$  & \multicolumn{2}{c|}{2}  & 2, 20 & \multicolumn{3}{c}{[2,100]\tnote{a}}    \\\midrule
    \makecell[cl]{Packet erasure rate \\ of device $k$, $p_k$}     & \makecell{$p_1=0.2$\\$p_2 = [0.1,0.6]$}   & \makecell{$p_1=0.5$\\$p_2 = [0.1,0.6]$}     &  [0.1,0.6], $\forall k \in \mathcal{K}$    &  0.25, $\forall k \in \mathcal{K}$  &  \makecell{0.05, if $1\le k \le \frac{K}{2}$\\0.45, if $\frac{K}{2}<k\le K$}  & $\frac{1}{2}\frac{k-1}{K-1}$ \\ \bottomrule
    \end{tabular}

    \begin{tablenotes}
     \item[a] $[a,b]$ means the value ranges from $a$ to $b$.
    \end{tablenotes}

    \end{threeparttable}
    \label{tab:parameters_for_6_3}
    \end{table*}
    
    \subsection{Observations from the actions obtained by the optimal and suboptimal policies}\label{opt_subopt_comparison}
    \begin{figure}
        \centering
        \includegraphics[width=\linewidth]{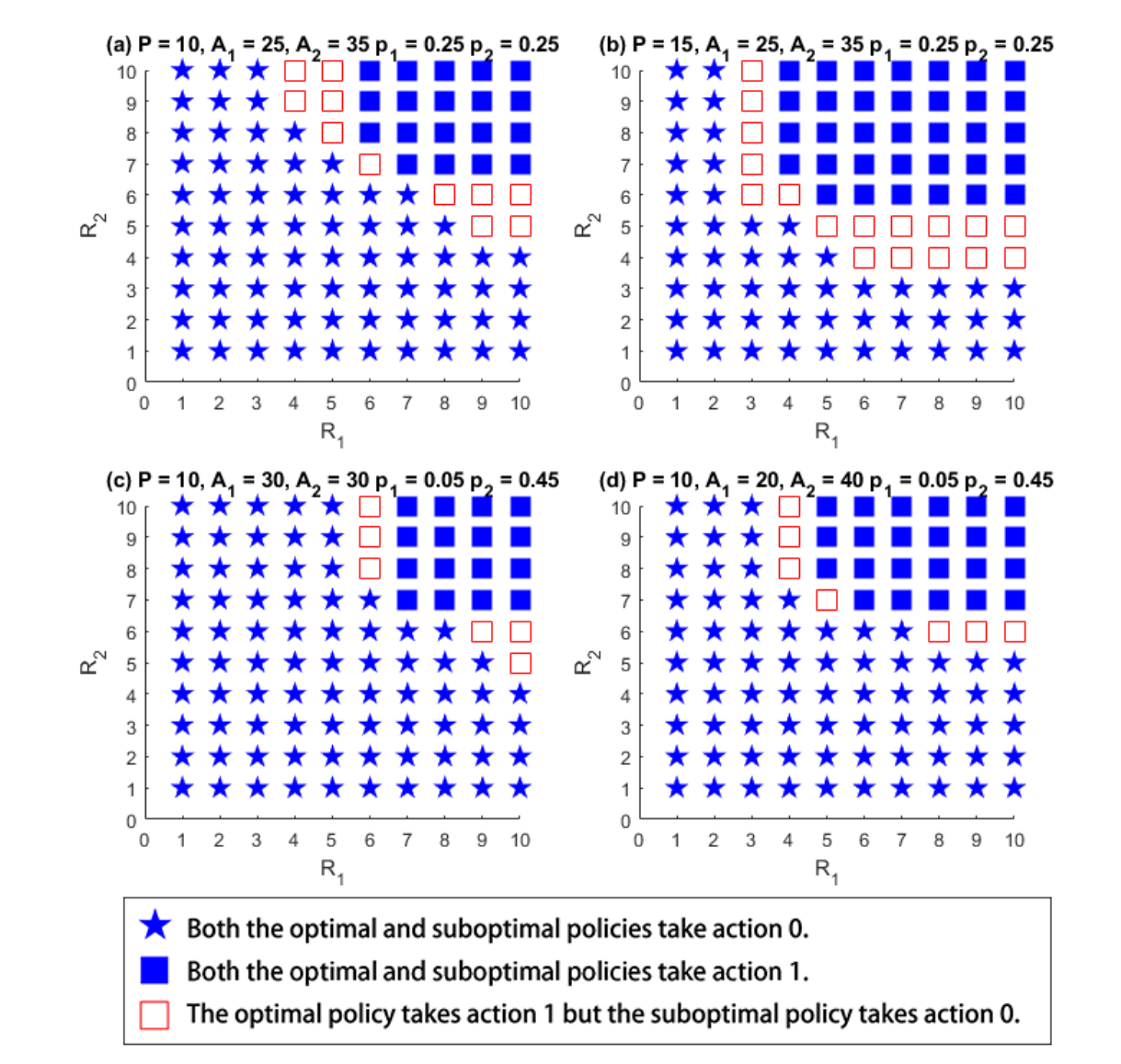}
        \caption{The actions taken by the suboptimal policy and optimal policy under different states and system settings. We fix $\xi = 10$, $\hat{A} = 40$ and $K=2$.}
        \label{fig:Structure_prop}
    \end{figure}
    
    \cref{fig:Structure_prop} illustrates the actions of the optimal policy and suboptimal policy under different states and system settings. The red and hollow symbols represent the states that the suboptimal solution will select a different action from the optimal one. First, we observe that the suboptimal policy has the same action as the optimal one in most states. This proves the validity of the low-complexity suboptimal solution. {It is worth noting that all the different actions stem from the suboptimal policy continuing to transmit the last status update. In contrast, the optimal policy decides to generate a new update.} This implies the suboptimal policy tends to overestimate the Q-factor under action 0. 
    
    Next, we focus on the structural properties—all four subfigures in \cref{fig:Structure_prop} indicate the structural property on $\mathbf{R}$ for both the optimal and suboptimal policies. If a strategy selects action~1 for a state, all actions to the top right of that state will also be 1. For the structural property on $P$, we shall focus on Figs.~\ref{fig:Structure_prop}(a) and~\ref{fig:Structure_prop}(b). {If the strategy has chosen to generate a new update on the state in \cref{fig:Structure_prop}(a), i.e., a smaller $P$, it will keep the same decision at the same position in \cref{fig:Structure_prop}(b). It means that if one state with a lower $P$ chooses to generate a new update, the other state will also follow suit, verifying the structural properties in \cref{thm:structure_prop_with_same_A} and \cref{cor:structural_prop_for_small_MDP}.} 
    
    Finally, we have some observations from the figure. Without the preemption, the source will wait for all devices to decode a status update so that it will not select to take action~1 unless $\mathbf{R} = \textbf{0}$. However, from the number of square symbols in the figure, preemption is essential for minimizing time-average AoI. In addition, we can infer some simple regularity from the actions. From Figs.~\ref{fig:Structure_prop}(a) and \ref{fig:Structure_prop}(b), when $p_k$ are the same, but $A_k$ are different, the source will deprioritize the device with lower $A_k$. For instance, the optimal policy will take action 1 when $\textbf{R}=[4,10]$ in \cref{fig:Structure_prop}(a), but will take action 0 when $\textbf{R}=[10,4]$. The same phenomenon can be seen in \cref{fig:Structure_prop}(c), where the states have the same $A_k$ while the packet erasure rates $p_k$ are distinct. The source will sacrifice the performance of a device with a lower packet erasure rate. In \cref{fig:Structure_prop}(d), device~1 has a lower age and packet erasure rate, making the phenomenon more significant. Consequently, the source shall deprioritize the device with lower AoI and better channel conditions to minimize the overall time-average AoI.

    {We conclude the main properties of the optimal and suboptimal policies: First, the suboptimal policy closely resembles the optimal policy in most states, validating the effectiveness of the low-complexity suboptimal solution. Second, the structural properties of both policies are evident in all parts of \cref{fig:Structure_prop}. Finally, the optimal policy may prioritize certain devices over others with lower AoI and better channel conditions to optimize the overall time-average AoI.}

    \subsection{Simulations for time-average AoI} \label{AoI_result}

    \begin{figure}
        \centering
        \includegraphics[width=\linewidth]{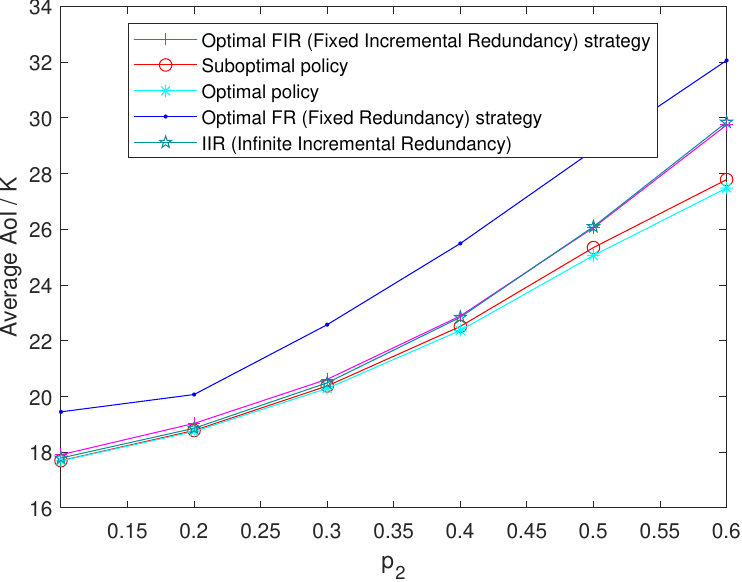}
        \caption{Average AoI per-device with respect to $p_2$ when $K=2$, $p_1 = 0.2$.}
        \label{fig:AoI_err_prob_1}
    \end{figure}

    \begin{figure}
        \centering
        \includegraphics[width=\linewidth]{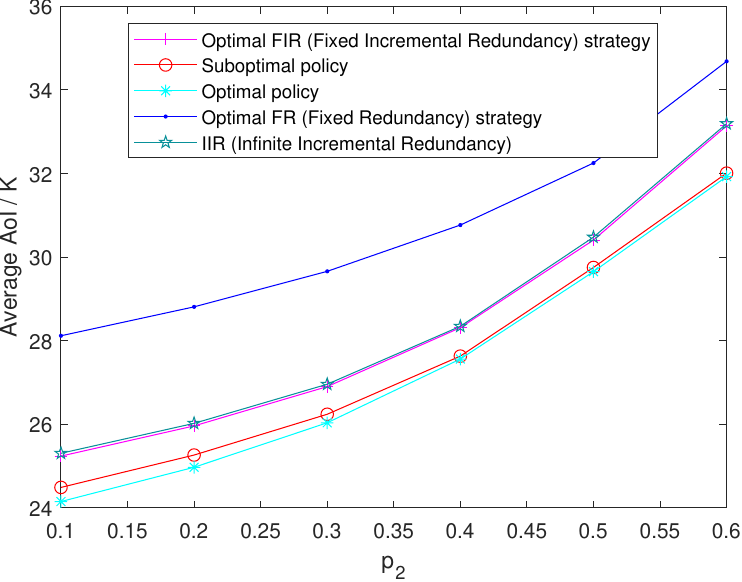}
        \caption{Average AoI per-device with respect to $p_2$ when $K=2$, $p_1 = 0.5$.}
        \label{fig:AoI_err_prob_2}
    \end{figure}

    \begin{figure}
        \centering
        \includegraphics[width=\linewidth]{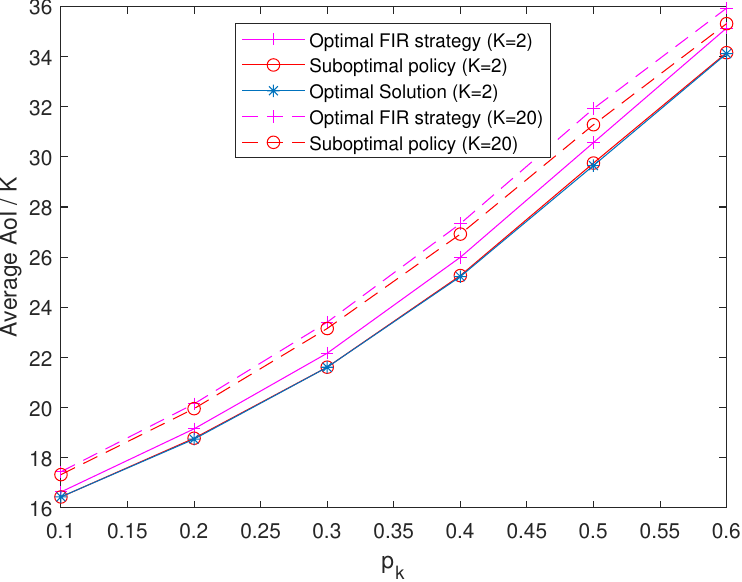}
        \caption{Average AoI per-device with respect to different packet erasure rates $p = p_k$ for all $k\in\mathcal{K}$. The dashed lines show age with $K=20$ and the solid lines show age with $K=2$.}
        \label{fig:AoI_err_prob_3}
    \end{figure}

    The first three figures fix the number of devices in the system while the packet erasure rate changes. \cref{fig:AoI_err_prob_1,fig:AoI_err_prob_2} illustrate the variation of AoI when we fix the packet erasure rate of one device, accompanied by the increase of the packet erasure rate of another device. The performance of our proposed suboptimal policy is almost the same as that of the optimal policy in these figures. Moreover, these two policies have a gap with other baseline policies, especially when the packet erasure rate increases. Among the baseline policies, the performance of the optimal FIR and IIR strategies is close, outperforming the optimal FR strategy. 
   
    \cref{fig:AoI_err_prob_3} assumes all devices have the same packet erasure rates and focus on two and twenty device cases, respectively. When there are two devices, the suboptimal policy exhibits almost the same AoI performance as the optimal policy, and both are better than the optimal FIR strategy. The suboptimal policy still outperforms the optimal FIR strategy when the number of devices reaches twenty. We conclude that our proposed low-complexity suboptimal policy almost overlaps the curve of the optimal policy {if the packet erasure rates are close}, and they both surpass the baseline strategies if we vary the packet erasure rates.

     \begin{figure}
        \centering
        \includegraphics[width=\linewidth]{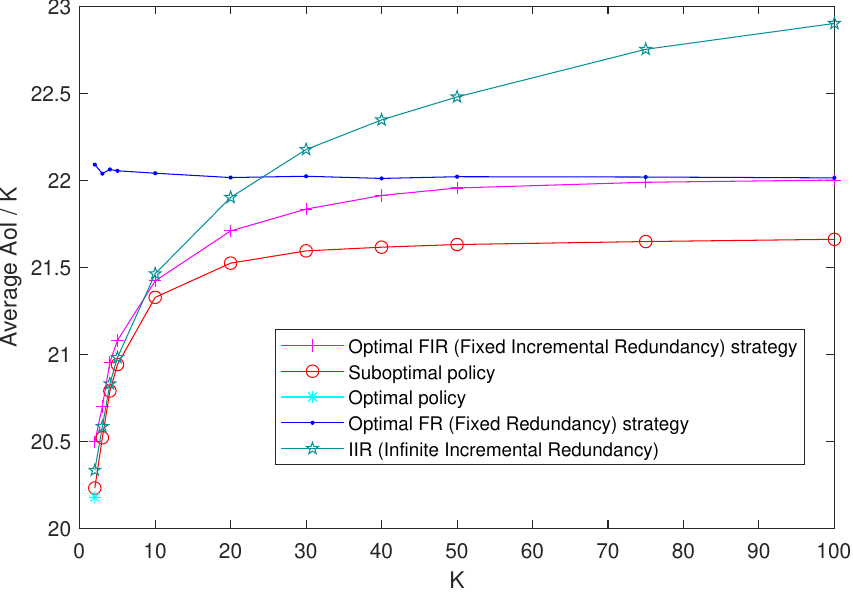}
        \caption{Average AoI per-device with respect to the number of devices $K$ with $p_k = 0.25$ for all $k\in\mathcal{K}$.}
        \label{fig:same_err}
    \end{figure}
    
    \begin{figure}
        \centering
        \includegraphics[width=\linewidth]{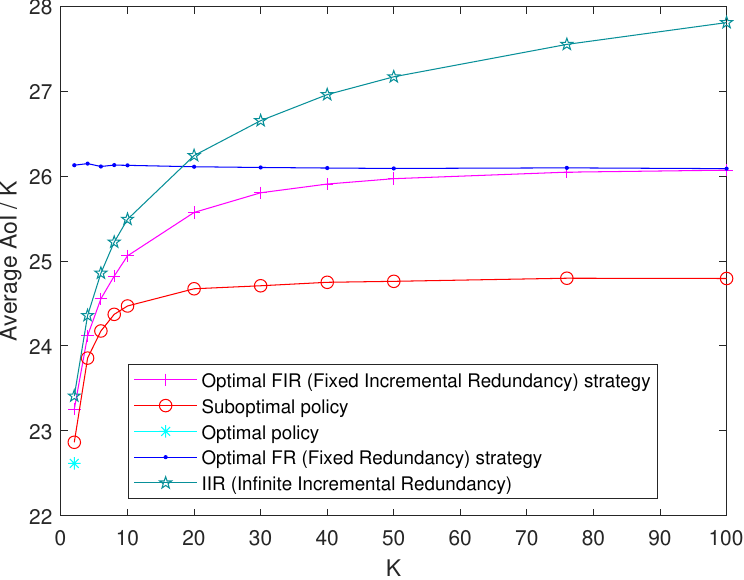}
        \caption{Average AoI per-device with respect to the number of devices $K$ with $p_k = 0.05$ if $1\le k \le \frac{K}{2}$ and $p_k = 0.45$ if $\frac{K}{2} < k \le K$ (All simulated $K$ in the above table are even numbers).}
        \label{fig:two_err}
    \end{figure}
    
    \begin{figure}
        \centering
        \includegraphics[width=\linewidth]{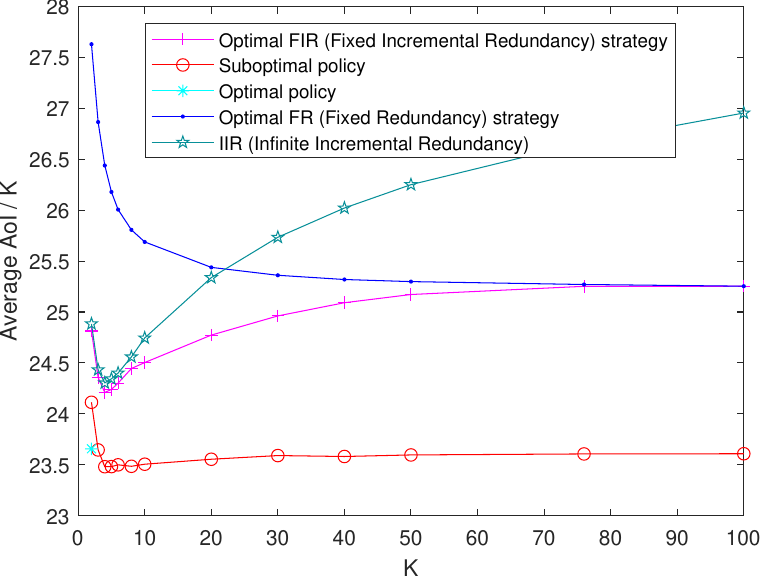}
        \caption{Average AoI per-device with respect to the number of devices $K$ with $p_k = \frac{1}{2}\frac{k-1}{K-1}$ for all $k\in\mathcal{K}$.}
        \label{fig:linear_err}
    \end{figure}
    
    \psfrag{E}{$\mathbbm{E}$}
    \begin{figure}
        \centering
        \includegraphics[width=\linewidth]{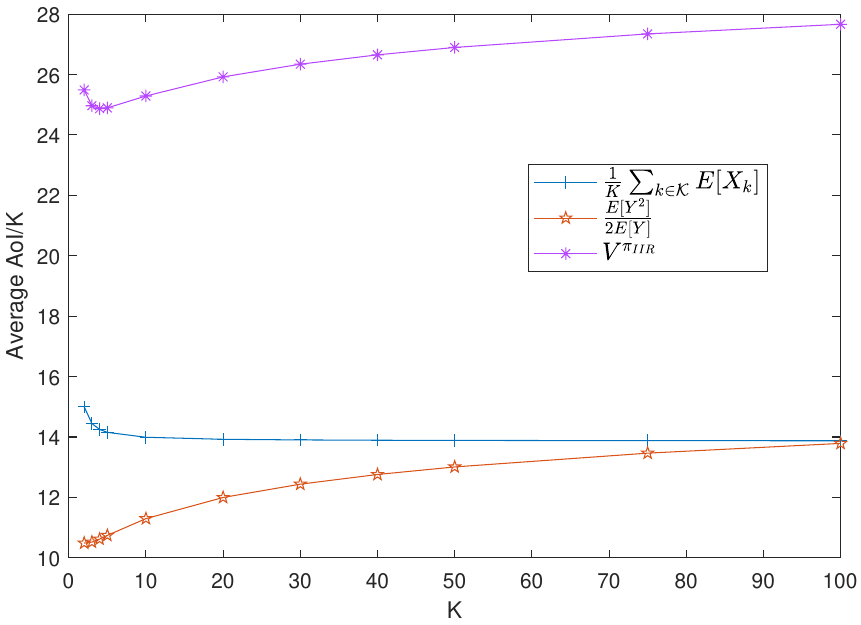}
        \caption{Theoretical Average AoI per-device under IIR strategy with respect to the number of devices $K$ with $p_k = \frac{1}{2}\frac{k-1}{K-1}$ for all $k\in\mathcal{K}$.}
        \label{fig:IIR analysis}
    \end{figure}
    \cref{fig:same_err,fig:two_err,fig:linear_err} investigate the average AoI when the number of devices changes with the same pattern of packet erasure rates. The mean of packet erasure rates is set to be 0.25. Due to the curse of dimensionality, the optimal solution is no longer available when the number of devices increases, so we only simulate it under the two-device case. 
    
    \cref{fig:same_err} specifies all devices' packet erasure rates to be {$0.25$}. Our proposed low-complexity suboptimal policy still outperforms other baseline policies. With the exception of the optimal FR strategy, AoI tends to increase as the number of devices increases. In addition, except for the IIR strategy, the AoI growth rate of other strategies decreases significantly when the number of devices is large. \cref{fig:two_err} divides devices into two groups, one with a high packet erasure rate of $0.45$, whereas the other group only has a low erasure rate {of} $0.05$. A similar phenomenon can be observed in this graph compared to that in \cref{fig:same_err}, while the gap between our proposed solution and the best baseline policy becomes greater. Furthermore, time-average AoI increases in all policies, implying the average AoI will increase if the packet erasure rates are not uniform. This shows that despite the same number of devices {and packet erasure rates' mean}, the time-average AoI can still be heavily influenced by the {distribution} of packet erasure rates.
    
    \cref{fig:linear_err} has the packet erasure rates linearly increase from 0 to 0.5 for each device, with the difference of $\frac{0.5}{K-1}$. Again, the performance of the suboptimal policy surpasses other baseline strategies. As the number of devices increases, the average AoI of the suboptimal strategy remains stable, and its gap with the optimal FIR strategy is the largest among the three cases. Interestingly, except for the optimal FR strategy, when the number of devices is small, the average AoI drops to a low point and slowly rises. Since it is hard to find the closed-form average AoI value of the suboptimal solution, we will explain it by the IIR strategy. It shows a similar trend to the suboptimal policy and has a closed-form solution. Based on (11) in \cite{yates2017timely}, we can derive the theoretical average AoI performance of IIR strategies $V^{\pi_{\text{IIR}}}$ without considering the upper limit of AoI. It can be written as 
    \begin{equation}
        V^{\pi_{\text{IIR}}} = \frac{1}{K}\sum_{k\in\mathcal{K}} \mathbbm{E}[X_k] + \frac{\mathbbm{E}[Y^2]}{2\mathbbm{E}[Y]},
    \end{equation}
    where $X_1, X_2, \dots$ are independent negative binomial (NB) ($\xi$,$1-p_k$) random variables, representing the number of transmissions required for each device to successfully decode an update. $Y = \max(X_1,X_2,\dots,X_K)$ is the maximum of all $X_k$. The first term $\frac{1}{K}\sum_{k\in\mathcal{K}} \mathbbm{E}[X_k]$ is the average expected number of packets for each device to receive $\xi$ packets. The second term $\frac{\mathbb{E}[Y^2]}{2\mathbb{E}[Y]}$ relates to $Y$. We expect $Y$ to be positively correlated with $K$, as we have to wait longer if more devices have to receive an update. In \cref{fig:IIR analysis}, we perform a simulation to explain the phenomenon further. The first term declines as the number of devices expands. By contrast, the second term increases gradually with the number of devices. We can observe that the first term shrinks significantly when the number of devices is small. This causes the total age to fall initially and goes up subsequently due to the increase in the second term\footnote{Since we have $\mathbbm{E}[X_k] = \frac{\xi}{1-p_k}$, the first term is equivalent to the multiplication of $\xi$ and the reciprocal of the harmonic mean of all $1-p_k$. Moreover, because $X_1, X_2, \dots$ are independent, the CDF of $Y$ can be denoted as $F_Y(y) = \prod_{k\in\mathcal{K}}F_k(y)$, where $F_k(x) = \sum_{i=\xi}^{x}{\binom{i-1}{\xi-1}}(1-p_k)^\xi p_k^{i-\xi}$ is the CDF of $X_k$. By utilizing $F_Y(y)$, the moments $\mathbb{E}[Y^2]$ and $\mathbb{E}[Y]$ can be deduced. Therefore, the simulation of \cref{fig:IIR analysis} is deterministic. Besides, $V^{\pi_{\text{IIR}}}$ will be slightly higher than the simulated result in \cref{fig:linear_err}. The reason is that $\hat{A}$ is infinite in \cref{fig:IIR analysis}, while we truncate the excessive age to a finite $\hat{A}$ in \cref{fig:linear_err}}.

    {\cref{fig:same_err,fig:two_err,fig:linear_err} also {reveal} an interesting phenomenon: when the number of devices is low, the IIR strategy exhibits superior performance in accurately capturing the retransmission timing. However, as the number of devices increases (with a threshold of 20 devices identified in our simulation), the IIR strategy, which waits for all devices to complete an update, shows significant shortcomings. In such scenarios, employing the Optimal FR strategy, which transmits a fixed number of packets without considering the reception status, efficiently addresses this issue. The Optimal FIR policy, integrating the advantages of both approaches, generally outperforms the former two under most conditions. Moreover, in scenarios where full feedback is available, the controller can further leverage this information to refine the transmission policy, achieving a superior average AoI compared to the initial three strategies, as demonstrated by our suboptimal policy.}
    
   In summary, the average AoI of our proposed suboptimal policy can approach the optimal one when the number of devices is low. The optimal solution is unavailable if the number of devices rises, but the suboptimal policy can still be used. In all settings, the suboptimal policy consistently outperforms other baseline policies. 

\section{Conclusions} \label{Section conclusion}
In future networks demanding real-time information and reduced bandwidth, multicast networks with age-aware features are emerging as a promising solution. For instance, in smart factories, a server multicasts environmental data to robots, and in wildfire detection, sensors adjust placements in real-time based on multicast directives. {We explored minimizing the time-average AoI in a multicast wireless sensor network using rateless codes, which eliminate retransmissions and save wireless resources.} Using MDP, we identified unique state transition properties and confirmed the MDP's communicating property, making the Bellman equation's solution feasible for an age-optimal policy. We also introduced a structure-aware RVIA for efficient policy solutions. For large-scale networks, we proposed a decentralized, low-complexity approach for a suboptimal policy.

Through numerical analysis, we demonstrated the effectiveness of both optimal and suboptimal policies and their structural properties. These policies often align in most states. {Following the proposed optimal scheduling policy, it is recommended that the source deprioritizes devices with lower AoI and better channel conditions to reduce the overall time-average AoI.} In simulations, our two policies consistently surpass baseline ones, with the suboptimal nearing the optimal performance. Despite identical average packet erasure rates among devices, the distribution still impacts AoI performance. AoI does not necessarily rise with more devices. In essence, our proposed solutions address the issue efficiently.

\appendices
\section{Proof of \texorpdfstring{\cref{lem: Existance of Bellman equation}}{Lemma~\ref{lem: Existance of Bellman equation}}}

We define the terms \textit{reach} and \textit{randomized stationary policy}.

\begin{definition}
    We say a state $X$ can \textit{reach} another state $X'$ if there exists a stationary policy $f$ such that for some $t$, we have
        \begin{equation}
        Pr[X(t)=X'|X(0)=X,f]>0
        \end{equation}
\end{definition}

\begin{definition}
A \textit{randomized} and stationary policy defined as $\hat{\pi}$ is a mapping from a set of states to a distribution on the action space $\mathcal{D} = \{0,1\}$. That is, $\hat{\pi}: \mathcal{X} \xrightarrow{} \mathscr{P}(\mathcal{D})$, where $\mathscr{P}(\mathcal{D})$ is a collection of probability distributions on subsets of $\mathcal{D}$. 
\end{definition}

If for every two states $X,X'\in\mathcal{X}$, we have $X$ can reach $X'$, then the MDP is communicating. Note that from \cite[Proposition 8.3.1]{Puterman}, the stationary policy $f$ is not required to be a deterministic policy. Therefore, we let $f$ be a stationary randomized policy of the MDP that chooses $D(t)=0$ and $D(t)=1$ with equal probability for all $t$, implying we may take action 0 or action 1 for every state. Before we continue to our proof, we first derive two useful lemmas by the age of information model.

\begin{lemma}\label{lem:reachability_1}
Given two states $X,X'\in\mathcal{X}$ where $X=(P,(A_k,R_k)_{k\in\mathcal{K}})$ and $X'=(1,\minhatA{A_k+1},R'_k)_{k\in\mathcal{K}})$, $X$ can reach $X'$.
\end{lemma}

\begin{proof}
The result is evident if we take action 1 on $X$, by \eqref{AoI_model_d1_P} and \eqref{AoI_model_d1_Ak}. 
\end{proof}

\begin{lemma}\label{lem:reachability_2}
Given two states $X,X'\in\mathcal{X}$ where $X=(1,(A_k,R_k)_{k\in\mathcal{K}})$ and $X'=(P',(A'_k,R'_k)_{k\in\mathcal{K}})$ with $P'>1$. If $A_k' = \minhatA{A_k + P' - 1}$ for those $k$ with $A'_k \neq P'$, and $R_k = \xi-1+\mathbbm{1}({R_k'=\xi}), \forall k\in\mathcal{K}$, then $X$ can reach $X'$. 
\end{lemma}

\begin{proof}
The following discussion focuses on the assumption the source takes action 0 starting from state $X$ for $P'-1$ times. Therefore, the limitation on $A_k$ and $R_k$ are both sufficient conditions of the lemma.

Under such assumption, the state $X'=(P',(A'_k,R_k')_{k\in\mathcal{K}})$ will have two possible values of $A'_k$ for each $k$.
If a device $k$ has successfully decoded the update, it will have $A'_k = P'$ by \eqref{AoI_model_d0_Xk}, and the initial value of $A_k$ is no longer important. On the other hand, if the device $k$ has not successfully decoded the update, its age will increase by $P'-1$ and thus $A_k' = \minhatA{A_k + P' - 1}$. Therefore, for the first constraint on $A'_k$, we know if $A'_k \neq P'$, then $A_k' = \minhatA{A_k + P' - 1}$ is the age at device $k$ after $P'-1$ times of transmission.

If $ R_k'=\xi$, it indicates the device $k$ has never succeeded in receiving any packet, including the first transmission. Therefore, we should set $R_k = \xi$. Otherwise, the device has at least succeeded in receiving a packet once, and we can assume it occurs at the first transmission and let $R_k=\xi-1$. Then, during the $P'-1$ times of transmission, we can choose $\xi-1-R'_k$ times of transmission to be successful because of the fact that $P'-1 \geq \xi-1-R'_k$ from \cref{prop:Rk_ge_xi_minus_P_if_P_le_xi} and $1-p_k>0$. 

Consequently, such constraint of $A_k$ and $R_k$ ensures that $X$ can reach $X'$ by $P'-1$ times of transmission with action 0 starting from state $X$.
\end{proof}

Combining the \cref{lem:reachability_1,lem:reachability_2}, we will have the following proposition.

\begin{proposition}\label{prop:combination_of_lemmas_to_show_accessibility}
Given two states $X,X'\in\mathcal{X}$ where $X=(P,(A_k,R_k)_{k\in\mathcal{K}})$ and $X'=(P',(A'_k,R'_k)_{k\in\mathcal{K}})$. If $A_k' = \minhatA{A_k + P'}$ for those $k$ with $A'_k \neq P'$, $X$ can reach $X'$.
\end{proposition}

Now, we come back to the original proof. Our target is to show for every state $X,X'\in\mathcal{X}$, $X$ can reach $X'$. Notice we have the observation that $R'_k$ will not influence the accessibility from \cref{prop:combination_of_lemmas_to_show_accessibility}. Therefore, we shall only focus on $P'$ and $A'_k$. For convenience, we denote $P'$ as $A_0'$. Without loss of generality, we assume $A_0'\le A'_1 \le A'_2 \le \cdots \le A'_K$. Besides, we define $\kappa$ as the smallest index such that $A'_k = \hat{A}$, where  
\begin{equation}
    \begin{aligned}
        \kappa=
            \begin{cases}
            \min \{k|A'_k = \hat{A},k\in\mathcal{K}\}&\text{if }\ A'_K=\hat{A}\\
            \infty & \text{otherwise.}
            \end{cases}
    \end{aligned}
\end{equation}

Then, we define the difference between each term as $e_k$ for all $1\le k\le K$, where
\begin{equation}
    \begin{aligned}
        e_k&=
            \begin{cases}
            \max\{A'_k-A'_{k-1},\xi\}&\text{if }\ k = \kappa\\
            A'_k-A'_{k-1}&\text{otherwise.}
            \end{cases}\\
    \end{aligned}
\end{equation}

Hereafter, the state $X'$ can be rewritten as $X'=(P',(\minhatA{P'+\sum_{j=1}^{k}e_j},R'_k)_{k\in\mathcal{K}})$. From \cref{prop:Rk_ge_xi_minus_P_if_P_le_xi}, we notice that $e_k = 0$ or $e_k\ge \xi$ if none of $A'_k$ equals to $\hat{A}$. By specifying the value of $e_\kappa$, we can guarantee there is no $0<e_k<\xi$. We define $\mathcal{I}_\xi$ as the set of index such that $e_k\ge \xi$, where $\mathcal{I}_\xi=\{i|e_i\ge \xi, i\in\mathcal{K}\}$, and $\bar{i}=\max\ \mathcal{I}_\xi$, $\ubar{i}=\min\ \mathcal{I}_\xi$. The biggest index that smaller than $i\in\mathcal{I}_\xi\setminus\{\ubar{i}\}$ is defined as $i^\star=\max\ \{j|j<i,j\in\mathcal{I}_\xi\}$.

Furthermore, for all $i\in\mathcal{I}_\xi$, we define its corresponding state $X_i = (P_i,(A_{k,i},R_{k,i})_{k\in\mathcal{K}}) =(e_i,(X_{k,i})_{k\in\mathcal{K}})$, where
\begin{equation}
    \begin{aligned}
        A_{k,i}&=
            \begin{cases}
            e_i&\text{if }k < i\\
            \sum_{j=i}^{k}e_j&\text{if }k\ge i\\
            \minhatA{\sum_{j=i}^{k}e_j}&\text{if }i=\ubar{i}, k=\kappa 
            \end{cases}\\
        R_{k,i} &= \xi\cdot\mathbbm{1}({A_{k,i} \neq e_i).}
    \end{aligned}
\end{equation}

We design these states so that we can prove $X_i$ can reach $X_{i^\star}$ iteratively. It is worth mentioning that only $A_{\kappa,\ubar{i}}$ may take the minimum with $\hat{A}$, as we will show in \eqref{42proof}. Because for all $k<\kappa$, $P'+\sum_{j=1}^{k}e_j<\hat{A}$. Besides, for $k=\kappa$, $e_\kappa$ may compensate at most $\xi$ for the difference. Note that if $\kappa \neq \infty$, $\bar{i} = \kappa$. By the facts that $e_{\ubar{i}}\ge \xi$ and $P'>0$, we have
\begin{equation}\label{42proof}
\begin{aligned}
    &P'+\sum_{j=1}^{\kappa}e_j-\hat{A} = P'+\sum_{j\in\mathcal{I}_\xi} e_j-\hat{A} \le \xi \\
\Leftrightarrow&\sum_{j\in\mathcal{I}_\xi\setminus\{\ubar{i}\}}e_j - \hat{A} \le \xi-P'-e_{\ubar{i}}<0\\
\Leftrightarrow&\sum_{j\in\mathcal{I}_\xi\setminus\{\ubar{i}\}}e_j < \hat{A}\Leftrightarrow A_{\kappa,i}<\hat{A}, \forall i\in\mathcal{I}_\xi\setminus\{\ubar{i}\}.
\end{aligned}
\end{equation}

Now, we will prove the three following things subsequently:
\begin{enumerate}
    \item \textit{$X$ can reach $X_{\bar{i}}$}: Since $e_k=0$ for all $k>\bar{i}$ by the definition of $\bar{i}$, we have $X_{\bar{i}} = (e_{\bar{i}},(e_{\bar{i}},0)_{k\in\mathcal{K}})$, which implying $A_{k,\bar{i}} = P_{\bar{i}}$ for all $k$. Afterward, from \cref{prop:combination_of_lemmas_to_show_accessibility}, every state can reach $X_{\bar{i}}$.
    
    \item \textit{$X_{\bar{i}}$ can reach $X_{\ubar{i}}$}: We will show that for all $i\in\mathcal{I}_\xi\setminus\{\ubar{i}\}$, $X_i$ can reach $X_{i^\star}$. From the definition of $X_{i^\star}$, we can find that for device $k<i$, $A_{k,i^\star}=P_{i^\star}$, so we do not have to concern the value of $A_{k,i}$. 
    Then, if $k\ge i$, since $A_{k,i}=\sum_{j=i}^{k}e_j$ if $i\in\mathcal{I}_\xi\setminus\{\ubar{i}\}$, $A_{k,i^\star} = \minhatA{\sum_{j=i^\star}^{k}e_j} = \minhatA{\sum_{j=i}^{k}e_j+\sum_{j=i^\star}^{i-1}e_j} = \minhatA{A_{k,i}+e_{i^\star}} = \minhatA{A_{k,i}+P_{i^\star}}$. Therefore, by the \cref{prop:combination_of_lemmas_to_show_accessibility}, $X_i$ can reach $X_{i^\star}$. By iteratively applying this result, we have $X_{\bar{i}}$ can reach $X_{\ubar{i}}$.
    
    \item \textit{$X_{\ubar{i}}$ can reach $X'$}: Note that $X'=(P',(\minhatA{P'+\sum_{j=1}^{k}e_j},R'_k)_{k\in\mathcal{K}})$. For those $k\ge\ubar{i}$, we have $A_k'>P'$. Then, $A_k' = \minhatA{P'+\sum_{j=\ubar{i}}^{k}e_j} =  \minhatA{P'+\minhatA{\sum_{j=\ubar{i}}^{k}e_j}} = \minhatA{P'+A_{k,\ubar{i}}}$. If $k<\ubar{i}$, $A_k'=P'$. Therefore, by \cref{prop:combination_of_lemmas_to_show_accessibility}, $X_{\ubar{i}}$ can reach $X'$.
\end{enumerate}

If $|\mathcal{I}_\xi|=1$, we shall skip the step 2 since $\bar{i}=\ubar{i}$. If $\mathcal{I}_\xi=\phi$, it means $P' = A'_k, \forall k\in\mathcal{K}$, so every state can reach $X'$ by \cref{prop:combination_of_lemmas_to_show_accessibility}. In summary, we can guarantee $X$ can always reach $X'$, and the proof is completed.

\section{Proof of \texorpdfstring{\cref{lem:increasing_relative_value_function}}{Lemma~\ref{lem:increasing_relative_value_function}}}
At the beginning, if $h_n(X^1)\le h_n(X^2)$ holds for all $n$, then by \eqref{Take limit to h_n}, we can take limit on both sides and thus $h(X^1) \le h(X^2)$. Therefore, we shall focus on the proof of $h_n(X^1)\le h_n(X^2)$ for all $n$ in the remaining part.

We will use the mathematical induction to prove the statement. Because we have initialized all $h_0(X)=0$, so it holds when $n=0$. Assume it holds for $n=n_0$ and by the condition $\mathbf{A}^1 \preceq \mathbf{A}^2$, $h_{n_0+1}(X^1)$ has the following relationship:
\begin{equation}
    \begin{aligned}
    h_{n_0+1}&(X^1)\\
    &= \sum_{k\in\mathcal{K}} A^1_k + \min_{d\in\mathcal{D}}\sum_{X^{1'}\in\mathcal{X}}Pr[X^{1'}|X^{1},d]h_{n_0}(X^{1'})\\
    &= \sum_{k\in\mathcal{K}} A^1_k + \sum_{X^{1'}\in\mathcal{X}}Pr[X^{1'}|X^{1},\pi^\star(X^1)]h_{n_0}(X^{1'})\\
    &\le \sum_{k\in\mathcal{K}} A^1_k + \sum_{X^{1'}\in\mathcal{X}}Pr[X^{1'}|X^{1},\pi^\star(X^2)]h_{n_0}(X^{1'})\\
    &\le \sum_{k\in\mathcal{K}} A^2_k + \sum_{X^{1'}\in\mathcal{X}}Pr[X^{1'}|X^{1},\pi^\star(X^2)]h_{n_0}(X^{1'})\\
    \end{aligned}
\end{equation}
The first inequality holds because $\pi^\star(X^1)$ is the optimal action for $X^1$, but $\pi^\star(X^2)$ is not necessarily. Similarly, $h_{n_0+1}(X^2)$ can be expressed as
\begin{equation}
\begin{aligned}
    h_{n_0+1}(&X^2)\\
    &=\sum_{k\in\mathcal{K}} A^2_k + \sum_{X^{2'}\in\mathcal{X}}Pr[X^{2'}|X^{2},\pi^\star(X^2)]h_{n_0}(X^{2'}).
\end{aligned}
\end{equation}

Therefore, to prove $h_{n_0+1}(X^1)\le h_{n_0+1}(X^2)$, it is sufficient to prove $\sum_{X^{1'}\in\mathcal{X}}Pr[X^{1'}|X^{1},\pi^\star(X^2)]h_{n_0}(X^{1'}) \le \sum_{X^{2'}\in\mathcal{X}}Pr[X^{2'}|X^{2},\pi^\star(X^2)]h_{n_0}(X^{2'})$. If $\pi^\star(X^2)=0$, then from \eqref{AoI_model_d0_P} to \eqref{AoI_model_d0_Xk}, we can verify that $P^{1'}\le P^{2'}$, $\mathbf{A}^{1'} \preceq \mathbf{A}^{2'}$ and $\mathbf{R}^{1'} \preceq \mathbf{R}^{2'}$ for all possible transitions\footnote{Since there are $K$ devices and each device may either succeed or fail to receive a packet, we have $2^K$ possible transitions after a decision.}. On the other hand, if $\pi^\star(X^2)=1$, we have $P^{1'} =  P^{2'}$, $\mathbf{A}^{1'} \preceq \mathbf{A}^{2'}$ and $\mathbf{R}^{1'} =  \mathbf{R}^{2'}$; as a result, regardless of $\pi^\star(X^2)$, one have  $h_{n_0}(X^{1'})\le h_{n_0}(X^{2'})$ according to the hypothesis of the induction. Hence, we conclude that $\sum_{X^{1'}\in\mathcal{X}}Pr[X^{1'}|X^{1},\pi^\star(X^2)]h_{n_0}(X^{1'}) \le \sum_{X^{2'}\in\mathcal{X}}Pr[X^{2'}|X^{2},\pi^\star(X^2)]h_{n_0}(X^{2'})$ holds, and it implies  $h_{n_0+1}(X^1)\le h_{n_0+1}(X^2)$. Consequently, by the mathematical induction, $h_n(X^1) \le h_n(X^2)$ holds for all $n$, and the proof of \cref{lem:increasing_relative_value_function} is completed.

\section{Proof of \texorpdfstring{\cref{thm:structure_prop_of_Q}}{Theorem~\ref{thm:structure_prop_of_Q}}}
We will start from the part of taking action 1. According to \eqref{Q-factor-iter}, it suffices to prove that the term  $\sum_{X^{1'}\in\mathcal{X}}Pr[X^{1'}|X^1,1]h_n(X^{1'})$ is equal to $\sum_{X^{2'}\in\mathcal{X}}Pr[X^{2'}|X^2,1]h_n(X^{2'})$ since $\mathbf{A}^1 = \mathbf{A}^2$.
According to the age of information model \eqref{AoI_model_d1_Ak} and \eqref{AoI_model_d1_Rk}, the two states will have the same transition for each device $k$ if their ages are the same. Therefore, $h_n(X^{1'}) = h_n(X^{2'})$ holds for all possible transitions, and the proof is completed.

For the part of taking action 0, by \eqref{Q-factor-iter}, it is equivalent to prove the inequality  $\sum_{X^{1'}\in\mathcal{X}}Pr[X^{1'}|X^{1},0]h_{n}(X^{1'}) \le \sum_{X^{2'}\in\mathcal{X}}Pr[X^{2'}|X^{2},0]h_{n}(X^{2'})$ instead. Under the premise $P^{1}\le P^{2}$, $\mathbf{A}^{1} = \mathbf{A}^{2}$ and $\mathbf{R}^{1} \preceq \mathbf{R}^{2'}$, one can check that from \eqref{AoI_model_d0_P} to \eqref{AoI_model_d0_Xk}, $P^{1'}\le P^{2'}$, $\mathbf{A}^{1'} \preceq \mathbf{A}^{2'}$ and $\mathbf{R}^{1'} \preceq \mathbf{R}^{2'}$ holds for all possible transitions if action 0 is taken. The age vector $\mathbf{A}^{1'}$ may become smaller because there may exist some $k$ who successfully received the packet, causing its $R_k^1$ to become 0, but $R_k^2$ does not. Thus $A_k^{1'}$ decreases to $P^{1'}$ by \eqref{AoI_model_d0_Xk}. Following \cref{lem:increasing_relative_value_function}, we deduce that $h_n(X^{1'}) \le h_n(X^{2'})$, and the proof is completed immediately. 

\section{Proof of \texorpdfstring{\cref{prop:R_is_xi_or_0_then_pi_is_1}}{Proposition~\ref{prop:R_is_xi_or_0_then_pi_is_1}}}
The proof will be divided into two parts. First, we will prove that if a state has $\mathbf{R} = \pmb{\xi}$, then $\pi^\star(X)=1$. According to the definition of the optimal action for a state in \eqref{definition of the optimal action}, our goal is to identify the action that minimizes the expected cost after its execution. For a state $X$ with $\mathbf{R}=\pmb{\xi}$, we denote the new state after taking action 0 as $X^{0'}$ and 1 as $X^{1'}$. We have $P^{0'}=\minhatA{P+1} > P^{1'}=1$ by \eqref{AoI_model_d0_P} and \eqref{AoI_model_d1_P}. Because $\mathbf{R}=\pmb{\xi}$, none of the devices can successfully decode an update and reduce its age at the next time slot, implying $\mathbf{A}^{0'} = \mathbf{A}^{1'}$ and $\mathbf{R}^{0'} = \mathbf{R}^{1'}$. 
After applying \cref{lem:increasing_relative_value_function}, we have $h(X^{0'}) \ge h(X^{1'})$. Combining the \eqref{definition of the optimal action}, we immediately obtain $\pi^\star(X) = 1$.

Let us move to the proof that if a state with $\mathbf{R} = \textbf{0}$, then $\pi^\star(X)=1$. 
We will again use mathematical induction to prove the statement. To make the proof more concise, we rewrite the narrative of the proof as the following lemma.
\begin{lemma}\label{lem:proof_in_prop_4}
For any $X^1 = (P^1,\mathbf{A}^1,\mathbf{R}^1), X^2 = (P^2,\mathbf{A}^2,\mathbf{R}^2) \in \mathcal{X}$ with $\mathbf{A}^1 = \mathbf{A}^2$ and $\mathbf{R}^1=\textbf{0}$, we have $h_n(X^1)\ge h_n(X^2)$ and $\pi_n^\star(X^1) = 1$ for all $n$. Hence, $h(X^1)\ge h(X^2)$ and $\pi^\star(X^1)=1$.
\end{lemma}
\begin{proof}
In the beginning, we define the optimal policy based on the relative value function at iteration $n$, $h_n(\cdot)$, as 
\begin{equation}\label{Optimal action under iteration n}
\begin{aligned}
        \pi^\star_{n}(X)  &= \arg\min_{d\in\mathcal{D}}Q_n(X,d)\\
    &= \arg\min_{d\in\mathcal{D}} \sum_{X'\in\mathcal{X}}Pr[X'|X,d]h_{n}(X')    
\end{aligned}
\end{equation}

Because we have initialized all $h_0(X)=0$, both statements are true for $n=0$. Assume $h_{n0}(X^1)\ge h_{n0}(X^2)$ is true for $n = n_0$. We first prove $\pi^\star_{n0}(X^1) = 1$ by discussing the transited states after taking different actions on $X^1$.

We denote the transited state after taking action $d$ for state $X^1$ as $X_d^{1'}$. If $d = 0$, since we have $\mathbf{R}^1 = \mathbf{0}$, it will follow \eqref{AoI_model_d0_Xk} and thus $X^{1'}_0 = (\minhatA{P^1+1},(\minhatA{A_k^1+1},0)_{k\in\mathcal{K}})$. On the other hand, if $d=1$, the state $X^{1'}_1$ becomes $(1,(\minhatA{A_k^1+1},R_k)_{k\in\mathcal{K}})$, where each $R_k$ is either $\xi$ or $\xi-1$ by \eqref{AoI_model_d1_Rk}. Note that $\mathbf{A}_0^{1'}$ $=$ $\mathbf{A}_1^{1'}$, and $\mathbf{R}_0^{1'} = \mathbf{0}$ under all possible transitions. Therefore, $h_{n0}(X^{1'}_0)\ge h_{n0}(X^{1'}_1)$ for all possible transition, and thus $\sum_{X^{1'}_0\in\mathcal{X}}Pr[X^{1'}_0|X^1,0]h_{n_0}(X^{1'}_0) \ge \sum_{X^{1'}_1\in\mathcal{X}}Pr[X^{1'}_1|X^1,1]h_{n_0}(X^{1'}_1)$, implying $\pi^\star_{n0}(X^1) = 1$ by the definition of \eqref{Optimal action under iteration n}.

Then, we will prove $h_{n_0+1}(X^1)\ge h_{n_0+1}(X^2)$ is true under the assumption  $h_{n_0}(X^1)\ge h_{n_0}(X^2)$. By observing \eqref{Q-factor-iter} and \eqref{relative-function-iter}, since $\mathbf{A}^1 = \mathbf{A}^2$ and $h_n(o)$ is prescribed, we only have to prove $\sum_{X^{2'}\in\mathcal{X}}Pr[X^{2'}|X^{2},\pi^\star_{n_0}(X^2)]h_{n_0}(X^{2'}) \le \sum_{X^{1'}\in\mathcal{X}}Pr[X^{1'}|X^{1},\pi^\star_{n_0}(X^1)]h_{n_0}(X^{1'})$:

\begin{equation}\label{lemma3_2}
\begin{aligned}
        &\sum_{X^{2'}\in\mathcal{X}}Pr[X^{2'}|X^{2},\pi^\star_{n_0}(X^2)]h_{n_0}(X^{2'})\\
    \le &\sum_{X^{2'}\in\mathcal{X}}Pr[X^{2'}|X^{2},1]h_{n_0}(X^{2'})\\
        = &\sum_{X^{1'}\in\mathcal{X}}Pr[X^{1'}|X^{1},1]h_{n_0}(X^{1'})\\
        = &\sum_{X^{1'}\in\mathcal{X}}Pr[X^{1'}|X^{1},\pi^\star_{n_0}(X^1)]h_{n_0}(X^{1'}).
\end{aligned}
\end{equation}
The inequality holds because $\pi^\star(X^2)$ is the optimal action for $X^2$. Thus, selecting action $\pi^\star_{n_0}(X^2)$ is always at least as good as selecting $1$. Furthermore, since $X^1$ and $X^2$ both have $\mathbf{A}^1=\mathbf{A}^2$, the first equality is validated by \cref{thm:structure_prop_of_Q}. Finally, given that $\mathbf{R}^1=\textbf{0}$, the second equality is confirmed by the proof from the previous part. As a result, we have proven that $h_{n_0+1}(X^1) \ge h_{n_0+1}(X^2)$ as indicated in \eqref{lemma3_2}. By mathematical induction the statement $h_n(X^1) \ge h_n(X^2)$ holds for every natural number $n$. It follows $\pi^\star_{n}(X^1) = 1$ holds for every natural number $n$, and thus  $\pi^\star(X^1) = 1$.
\end{proof}

In \cref{lem:proof_in_prop_4}, if $X^2 = X^1$, $\pi^\star(X^1)=1$ still holds. Accordingly, we conclude $\pi^\star(X) = 1$ if $\mathbf{R}=\pmb{\xi}$ or $\mathbf{R}=\mathbf{0}$, by the first paragraph and \cref{lem:proof_in_prop_4}, respectively.

\section*{References}
\bibliography{references}

\begin{thebibliography}{10}
\providecommand{\url}[1]{#1}
\csname url@samestyle\endcsname
\providecommand{\newblock}{\relax}
\providecommand{\bibinfo}[2]{#2}
\providecommand{\BIBentrySTDinterwordspacing}{\spaceskip=0pt\relax}
\providecommand{\BIBentryALTinterwordstretchfactor}{4}
\providecommand{\BIBentryALTinterwordspacing}{\spaceskip=\fontdimen2\font plus
\BIBentryALTinterwordstretchfactor\fontdimen3\font minus \fontdimen4\font\relax}
\providecommand{\BIBforeignlanguage}[2]{{%
\expandafter\ifx\csname l@#1\endcsname\relax
\typeout{** WARNING: IEEEtran.bst: No hyphenation pattern has been}%
\typeout{** loaded for the language `#1'. Using the pattern for}%
\typeout{** the default language instead.}%
\else
\language=\csname l@#1\endcsname
\fi
#2}}
\providecommand{\BIBdecl}{\relax}
\BIBdecl

\bibitem{Chukhno2023Approaching6G}
N.~Chukhno, O.~Chukhno, S.~Pizzi, A.~Molinaro, A.~Iera, and G.~Araniti, ``Approaching 6g use case requirements with multicasting,'' \emph{IEEE Communications Magazine}, vol.~61, no.~5, pp. 144--150, 2023.

\bibitem{Kaul2011Minimizing}
S.~Kaul, M.~Gruteser, V.~Rai, and J.~Kenney, ``Minimizing age of information in vehicular networks,'' in \emph{2011 8th Annual IEEE Communications Society Conference on Sensor, Mesh and Ad Hoc Communications and Networks}, 2011, pp. 350--358.

\bibitem{Kaul2011Onpiggybacking}
S.~Kaul, R.~Yates, and M.~Gruteser, ``On piggybacking in vehicular networks,'' in \emph{2011 IEEE Global Telecommunications Conference (GLOBECOM)}, 2011, pp. 1--5.

\bibitem{Kaul2012Realtime}
------, ``Real-time status: How often should one update?'' in \emph{2012 Proceedings IEEE INFOCOM}, 2012, pp. 2731--2735.

\bibitem{3gpp.36.331}
3GPP, ``{Evolved Universal Terrestrial Radio Access (E-UTRA); Radio Resource Control (RRC); Protocol specification},'' {3rd Generation Partnership Project (3GPP)}, Technical Specification (TS) 26.346, 08 2022, version 17.0.0.

\bibitem{3gpp.23.247}
------, ``{Architectural enhancements for 5G multicast-broadcast services },'' {3rd Generation Partnership Project (3GPP)}, Technical Specification (TS) 23.247, 04 2023, version 17.6.0.

\bibitem{3gpp.23.774}
------, ``{Study on mission critical services over 5G multicast broadcast system},'' {3rd Generation Partnership Project (3GPP)}, Technical Report (TR) 23.774, 03 2020, version 17.0.0.

\bibitem{3gpp.26.502}
------, ``{5G multicast-broadcast services; User service architecture},'' {3rd Generation Partnership Project (3GPP)}, Technical Specification (TS) 26.502, 04 2023, version 17.4.0.

\bibitem{7866892}
A.~Rajanna and M.~Haenggi, ``Enhanced cellular coverage and throughput using rateless codes,'' \emph{IEEE Transactions on Communications}, vol.~65, no.~5, pp. 1899--1912, 2017.

\bibitem{Luby2002LTcodes}
M.~Luby, ``Lt codes,'' in \emph{The 43rd Annual IEEE Symposium on Foundations of Computer Science, 2002. Proceedings.}, 2002, pp. 271--280.

\bibitem{Shokrollahi2006Raptorcode}
A.~Shokrollahi, ``Raptor codes,'' \emph{IEEE Transactions on Information Theory}, vol.~52, no.~6, pp. 2551--2567, 2006.

\bibitem{rfc6330}
\BIBentryALTinterwordspacing
L.~Minder, A.~Shokrollahi, M.~Watson, M.~Luby, and T.~Stockhammer, ``{RaptorQ Forward Error Correction Scheme for Object Delivery},'' RFC 6330, Aug. 2011. [Online]. Available: \url{https://www.rfc-editor.org/info/rfc6330}
\BIBentrySTDinterwordspacing

\bibitem{Raptor2011}
\BIBentryALTinterwordspacing
A.~Shokrollahi and M.~Luby, ``Raptor codes,'' \emph{Foundations and Trends® in Communications and Information Theory}, vol.~6, no. 3–4, pp. 213--322, 2011. [Online]. Available: \url{http://dx.doi.org/10.1561/0100000060}
\BIBentrySTDinterwordspacing

\bibitem{yates2017timely}
R.~D. Yates, E.~Najm, E.~Soljanin, and J.~Zhong, ``Timely updates over an erasure channel,'' in \emph{2017 IEEE International Symposium on Information Theory (ISIT)}.\hskip 1em plus 0.5em minus 0.4em\relax IEEE, 2017, pp. 316--320.

\bibitem{AoI_survey}
R.~D. Yates, Y.~Sun, D.~R. Brown, S.~K. Kaul, E.~Modiano, and S.~Ulukus, ``Age of information: An introduction and survey,'' \emph{IEEE Journal on Selected Areas in Communications}, vol.~39, no.~5, pp. 1183--1210, 2021.

\bibitem{Ostman2019fading}
J.~Östman, R.~Devassy, G.~Durisi, and E.~Uysal, ``Peak-age violation guarantees for the transmission of short packets over fading channels,'' in \emph{IEEE INFOCOM 2019 - IEEE Conference on Computer Communications Workshops (INFOCOM WKSHPS)}, 2019, pp. 109--114.

\bibitem{Talak2021multihop}
V.~Tripathi, R.~Talak, and E.~Modiano, ``Information freshness in multi-hop wireless networks,'' \emph{IEEE/ACM Transactions on Networking}, 2022.

\bibitem{Yates2017multiaccess}
R.~D. Yates and S.~K. Kaul, ``Status updates over unreliable multiaccess channels,'' in \emph{2017 IEEE International Symposium on Information Theory (ISIT)}, 2017, pp. 331--335.

\bibitem{Zhang2021pricing}
M.~Zhang, A.~Arafa, J.~Huang, and H.~V. Poor, ``Pricing fresh data,'' \emph{IEEE Journal on Selected Areas in Communications}, vol.~39, no.~5, pp. 1211--1225, 2021.

\bibitem{Zhang2022AoIvehicular}
S.~Zhang, J.~Li, H.~Luo, J.~Gao, L.~Zhao, and X.~Sherman~Shen, ``Low-latency and fresh content provision in information-centric vehicular networks,'' \emph{IEEE Transactions on Mobile Computing}, vol.~21, no.~5, pp. 1723--1738, 2022.

\bibitem{Kadota2018broadcast}
I.~Kadota, A.~Sinha, E.~Uysal-Biyikoglu, R.~Singh, and E.~Modiano, ``Scheduling policies for minimizing age of information in broadcast wireless networks,'' \emph{IEEE/ACM Transactions on Networking}, vol.~26, no.~6, pp. 2637--2650, 2018.

\bibitem{Hsu2020scheduling}
Y.-P. Hsu, E.~Modiano, and L.~Duan, ``Scheduling algorithms for minimizing age of information in wireless broadcast networks with random arrivals,'' \emph{IEEE Transactions on Mobile Computing}, vol.~19, no.~12, pp. 2903--2915, 2020.

\bibitem{Kadota2021MinimizingStochastic}
I.~Kadota and E.~Modiano, ``Minimizing the age of information in wireless networks with stochastic arrivals,'' \emph{IEEE Transactions on Mobile Computing}, vol.~20, no.~3, pp. 1173--1185, 2021.

\bibitem{Ceran2021ARL}
E.~T. Ceran, D.~G{\"u}nd{\"u}z, and A.~Gy{\"o}rgy, ``A reinforcement learning approach to age of information in multi-user networks with {HARQ},'' \emph{IEEE Journal on Selected Areas in Communications}, vol.~39, pp. 1412--1426, 2021.

\bibitem{10198349}
T.~Zhang, S.~Chen, Z.~Chen, Z.~Tian, Y.~Jia, M.~Wang, and D.~O. Wu, ``{AoI} and {PAoI} in the {IoT}-based multisource status update system: Violation probabilities and optimal arrival rate allocation,'' \emph{IEEE Internet of Things Journal}, vol.~10, no.~23, pp. 20\,617--20\,632, 2023.

\bibitem{10078176}
Z.~Chen, D.~Deng, H.~H. Yang, N.~Pappas, L.~Hu, Y.~Jia, M.~Wang, and T.~Q.~S. Quek, ``Analysis of {Age of Information} in dual updating systems,'' \emph{IEEE Transactions on Wireless Communications}, vol.~22, no.~11, pp. 8003--8019, 2023.

\bibitem{10278911}
K.~Lang, Z.~Chen, N.~Pappas, H.~H. Yang, Y.~Jia, and T.~Q.~S. Quek, ``The effect of device redundancy in timeliness of information,'' in \emph{ICC 2023 - IEEE International Conference on Communications}, 2023, pp. 1694--1699.

\bibitem{Tang2023BroadcastOrUnicast}
Z.~Tang, N.~Yang, P.~Sadeghi, and X.~Zhou, ``Age of information in downlink systems: Broadcast or unicast transmission?'' \emph{IEEE Journal on Selected Areas in Communications}, vol.~41, no.~7, pp. 2057--2070, 2023.

\bibitem{zhong2017status}
J.~Zhong, E.~Soljanin, and R.~D. Yates, ``Status updates through multicast networks,'' in \emph{2017 55th Annual Allerton Conference on Communication, Control, and Computing (Allerton)}.\hskip 1em plus 0.5em minus 0.4em\relax IEEE, 2017, pp. 463--469.

\bibitem{zhong2018multicast}
J.~Zhong, R.~D. Yates, and E.~Soljanin, ``Multicast with prioritized delivery: How fresh is your data?'' in \emph{2018 IEEE 19th International Workshop on Signal Processing Advances in Wireless Communications (SPAWC)}.\hskip 1em plus 0.5em minus 0.4em\relax IEEE, 2018, pp. 1--5.

\bibitem{buyukates2019age}
B.~Buyukates, A.~Soysal, and S.~Ulukus, ``Age of information in multihop multicast networks,'' \emph{Journal of Communications and Networks}, vol.~21, no.~3, pp. 256--267, 2019.

\bibitem{chen2020optimizing}
Y.~Chen, H.~Tang, and J.~Wang, ``Optimizing age of information in multicast unilateral networks,'' in \emph{2020 IEEE International Symposium on Broadband Multimedia Systems and Broadcasting (BMSB)}.\hskip 1em plus 0.5em minus 0.4em\relax IEEE, 2020, pp. 1--4.

\bibitem{Li2020MulticastDeadline}
J.~Li, Y.~Zhou, and H.~Chen, ``Age of information for multicast transmission with fixed and random deadlines in {IoT} systems,'' \emph{IEEE Internet of Things Journal}, vol.~7, no.~9, pp. 8178--8191, 2020.

\bibitem{hsu2021efficient}
Y.-P. Hsu, ``Efficient broadcast for timely updates in mobile networks,'' \emph{IEEE Communications Letters}, vol.~25, no.~6, pp. 1969--1973, 2021.

\bibitem{nath2018optimum}
S.~Nath, J.~Wu, and J.~Yang, ``Optimum energy efficiency and age-of-information tradeoff in multicast scheduling,'' in \emph{2018 IEEE International Conference on Communications (ICC)}.\hskip 1em plus 0.5em minus 0.4em\relax IEEE, 2018, pp. 1--6.

\bibitem{xie2020age}
M.~Xie, J.~Gong, and X.~Ma, ``Age-energy tradeoff of short packet based transmissions in multicast networks with {ARQ},'' in \emph{2020 IEEE 91st Vehicular Technology Conference (VTC2020-Spring)}.\hskip 1em plus 0.5em minus 0.4em\relax IEEE, 2020, pp. 1--5.

\bibitem{Wei2020Ontheage}
H.~Wei and N.~Deng, ``On the age of information in wireless networks using rateless codes,'' \emph{IEEE Access}, vol.~8, pp. 173\,147--173\,157, 2020.

\bibitem{Meng2021Spinal}
S.~Meng, S.~Wu, A.~Li, J.~Jiao, N.~Zhang, and Q.~Zhang, ``Analysis and optimization of spinal codes over the bsc: from the aoi perspective,'' in \emph{2021 IEEE International Conference on Communications Workshops (ICC Workshops)}, 2021, pp. 1--6.

\bibitem{Wang2019whento}
B.~Wang, S.~Feng, and J.~Yang, ``When to preempt? age of information minimization under link capacity constraint,'' \emph{Journal of Communications and Networks}, vol.~21, no.~3, pp. 220--232, 2019.

\bibitem{Wang2022AgeOptimalTP}
Y.~Wang, S.~Wu, J.~Jiao, W.~Wu, Y.~Wang, and Q.~Zhang, ``Age-optimal transmission policy with {HARQ} for freshness-critical vehicular status updates in space–air–ground-integrated networks,'' \emph{IEEE Internet of Things Journal}, vol.~9, no.~8, pp. 5719--5729, 2022.

\bibitem{Zhou2020minimum}
B.~Zhou and W.~Saad, ``Minimum age of information in the internet of things with non-uniform status packet sizes,'' \emph{IEEE Transactions on Wireless Communications}, vol.~19, no.~3, pp. 1933--1947, 2020.

\bibitem{Xie2021online}
X.~Xie, H.~Wang, L.~Yu, and M.~Weng, ``Online algorithms for optimizing age of information in the iot systems with multi-slot status delivery,'' \emph{IEEE Wireless Communications Letters}, vol.~10, no.~5, pp. 971--975, 2021.

\bibitem{Feng2019Ageoptimal}
S.~Feng and J.~Yang, ``Age-optimal transmission of rateless codes in an erasure channel,'' in \emph{2019 IEEE International Conference on Communications (ICC)}, 2019, pp. 1--6.

\bibitem{Gindullina2021AoIupper_stale}
E.~Gindullina, L.~Badia, and D.~Gündüz, ``Age-of-information with information source diversity in an energy harvesting system,'' \emph{IEEE Transactions on Green Communications and Networking}, vol.~5, no.~3, pp. 1529--1540, 2021.

\bibitem{Leng2019AoIupper_stale}
S.~Leng and A.~Yener, ``Age of information minimization for an energy harvesting cognitive radio,'' \emph{IEEE Transactions on Cognitive Communications and Networking}, vol.~5, no.~2, pp. 427--439, 2019.

\bibitem{Puterman}
M.~L. Puterman, \emph{Markov Decision Processes: Discrete Stochastic Dynamic Programming}, 1st~ed.\hskip 1em plus 0.5em minus 0.4em\relax USA: John Wiley \& Sons, Inc., 1994.

\bibitem{Bertsekas}
D.~P. Bertsekas, \emph{Dynamic Programming and Optimal Control}, 3rd~ed.\hskip 1em plus 0.5em minus 0.4em\relax Athena Scientific, 2007, vol.~2.

\bibitem{powell2007approximate}
W.~B. Powell, \emph{Approximate Dynamic Programming: Solving the curses of dimensionality}.\hskip 1em plus 0.5em minus 0.4em\relax John Wiley \& Sons, 2007, vol. 703.

\bibitem{bedewy2021optimalSampling}
A.~M. Bedewy, Y.~Sun, S.~Kompella, and N.~B. Shroff, ``Optimal sampling and scheduling for timely status updates in multi-source networks,'' \emph{IEEE Transactions on Information Theory}, vol.~67, no.~6, pp. 4019--4034, 2021.

\end{thebibliography}

\begin{IEEEbiography}[{\includegraphics[width=1in,height=1.25in,clip,keepaspectratio]{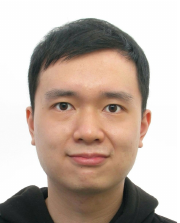}}]{Hung-Chun Lin} received the B.S. degree in electrical engineering from National Taiwan University, Taipei, Taiwan, in 2021. He is currently pursuing his M.S. degree in communication engineering at Nation Taiwan University, Taipei, Taiwan.

He received Academic Excellence Award from National Taiwan University in 2020. His research interests include wireless networks, information freshness, reinforcement learning and power-saving mechanisms. 
\end{IEEEbiography}

\begin{IEEEbiography}[{\includegraphics[width=1in,height=1.25in,clip,keepaspectratio]{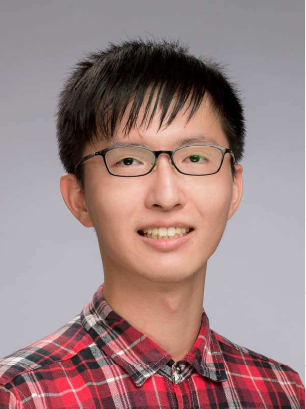}}]{Kuang-Hsun Lin} (Member, IEEE) obtained his B.S. degree in Electrical Engineering from National Taiwan University in 2015 and his Ph.D. degree in Communication Engineering from GICE at National Taiwan University in 2022. Since 2015, he has been working under the supervision of Prof. H.-Y. Wei at the Wireless and Mobile Networking Laboratory. He completed summer internships with MediaTek in 2015 and 2018. His research interests include wireless mobile networks, power-saving mechanisms for devices and networks, mobility management, and MAC protocol design. Currently, he is a Postdoctoral Researcher at GICE, National Taiwan University, and also serves as a Postdoctoral Research Assistant for Taiwan's NSTC 6G Program Office.
\end{IEEEbiography}

\vfill

\begin{IEEEbiography}[{\includegraphics[width=1in,height=1.25in,clip,keepaspectratio]{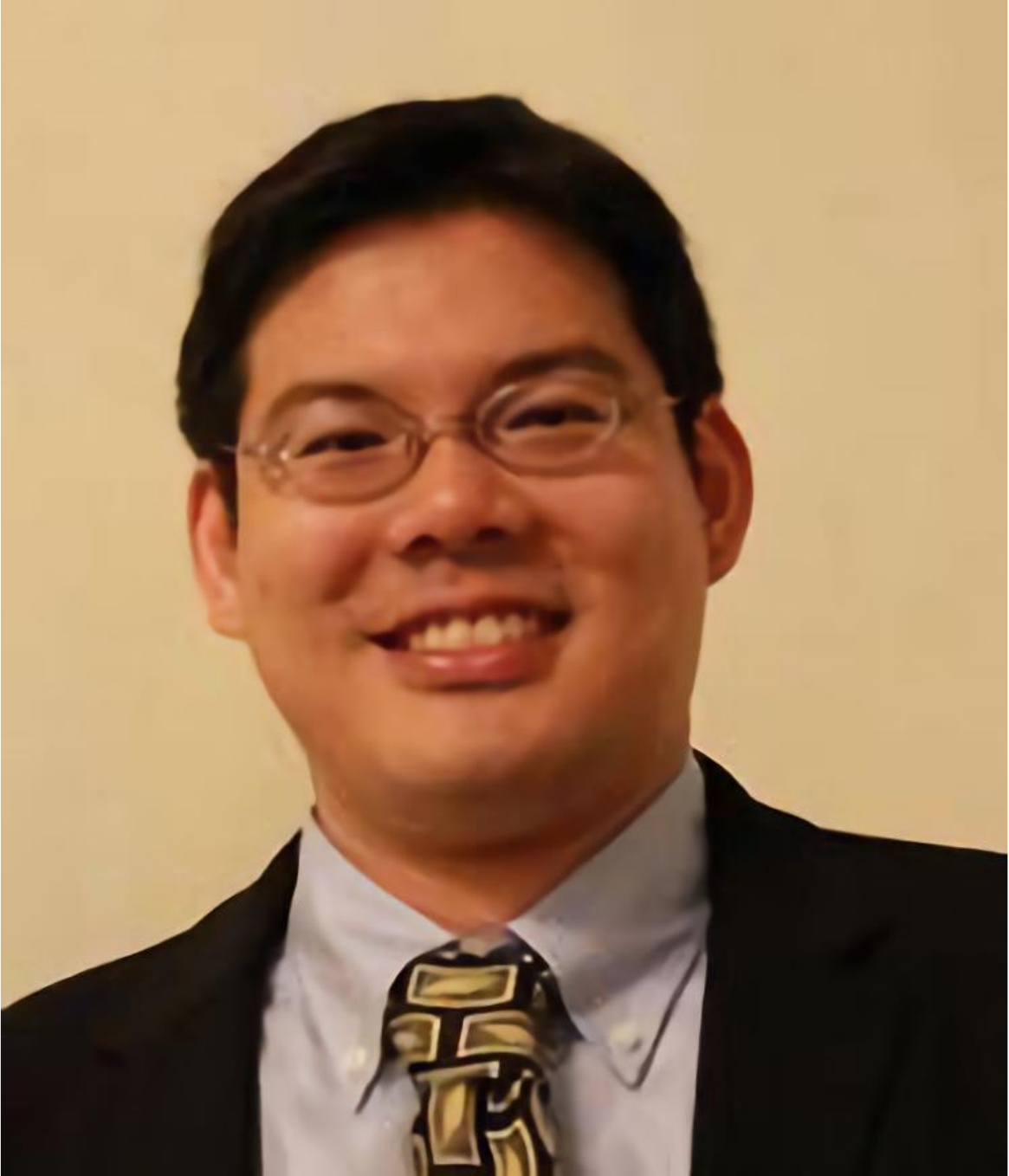}}]{Hung-Yu Wei} (Senior Member, IEEE) is a Professor in Department of Electrical Engineering and Graduate Institute of Communications Engineering, National Taiwan University. He served as Interim Department Chair and Associate Chair in Department of Electrical Engineering. He received the B.S. degree in electrical engineering from National Taiwan University in 1999. He received the M.S. and the Ph.D. degree in electrical engineering from Columbia University in 2001 and 2005 respectively. He was a summer intern at Telcordia Applied Research in 2000 and 2001. He was with NEC Labs America from 2003 to 2005. He joined Department of Electrical Engineering at the National Taiwan University in July 2005. His research interests include next-generation wireless networks, IoT, fog/edge computing, cross-layer design for wireless multimedia, and game theoretic models for communications networks.
	
Dr. Wei received NTU Excellent Teaching Award in 2008 and 2018. He also received ``Recruiting Outstanding Young Scholar Award'' from the Foundation for the Advancement of Outstanding Scholarship in 2006, K.~T.~Li Young Researcher Award from ACM Taipei/Taiwan Chapter and The Institute of Information and Computing Machinery in 2012, Excellent Young Engineer Award from the Chinese Institute of Electrical Engineering in 2014, Wu~Ta~You Memorial Award from MOST in 2015, and Outstanding Research Award from MOST in 2020. He has been actively participating in NGMN, IEEE 802.16, 3GPP, IEEE P1934, and IEEE P1935 standardization. He serves as Secretary for IEEE ComSoC Fog/Edge Industry Community. He serves as Chair of IEEE P1935 working group for edge/fog management and orchestration standard and Chair of IEEE P1934 Working Group for edge/fog computing and networking architecture. He is an Associate Editor for \textit{IEEE IoT magazine}, and was an Associate Editor for \textsc{IEEE IoT Journal} and \textsc{IEEE System journal}. He is an IEEE certified Wireless Communications Professional. He was the Chair of IEEE VTS Taipei Chapter during 2016--2017.
\end{IEEEbiography}

\vfill

\end{document}